\documentclass[twocolumn]{aastex62}
\usepackage{graphicx}
\usepackage{rotating}
\usepackage{latexsym}
\usepackage{lineno}
\usepackage{amsmath}

\DeclareRobustCommand{\ion}[2]{%
\relax\ifmmode
\ifx\testbx\f@series
{\mathbf{#1\,\mathsc{#2}}}\else
{\mathrm{#1\,\mathsc{#2}}}\fi
\else\textup{#1\,{\mdseries\textsc{#2}}}%
\fi}

\shorttitle{$\textit{WISE}$ study of G23}
\shortauthors{H. F. M. Yao et al.}

\begin{document}

\title{Connecting MeerKAT radio continuum properties to GAMA optical emission-line and $\textit{WISE}$ mid-infrared activity}

\correspondingauthor{Harris Yao Fortune Marc}
\email{yao.fortune@gmail.com}

\author[0000-0001-8459-4034]{H. F. M. Yao}
\affil{Department of Physics and Astronomy, University of the Western Cape, Robert Sobukwe Road, Bellville, 7535, Republic of South Africa}
\affil{Department of Physics, University of  Felix  Houphouet Boigny, 01 BP V 34 Abidjan 01 Cote d'Ivoire}

\author{M. E. Cluver}
\affiliation{Centre for Astrophysics and Supercomputing, Swinburne University of Technology, John Street, Hawthorn, 3122, Australia}
\affiliation{Department of Physics and Astronomy, University of the Western Cape, Robert Sobukwe Road, Bellville, 7535, Republic of South Africa}

\author{T. H. Jarrett}
\affil{Department of Astronomy, University of Cape Town, Private Bag X3, Rondebosch, 7701, South Africa}

\author{G. I. G. J\'ozsa}
\affiliation{Max-Planck-Institut f\"ur Radioastronomie, Radioobservatorium Effelsberg, Max-Planck-Stra{\ss}e 28, D-53902 Bad M\"unstereifel, Germany; gjozsa@mpifr-bonn.mpg.de}
\affiliation{Department of Physics and Electronics, Rhodes University, PO Box 94, Makhanda, 6140, South Africa}
\affiliation{South African Radio Astronomy Observatory, 2 Fir Street, Black River Park, Observatory, Cape Town, 7925, South Africa}

\author{M. G. Santos}
\affiliation{Department of Physics and Astronomy, University of the Western Cape, Robert Sobukwe Road, Bellville, 7535, Republic of South Africa}

\author{L. Marchetti}
\affiliation{Department of Physics and Astronomy, University of the Western Cape, Robert Sobukwe Road, Bellville, 7535, Republic of South Africa}
\affiliation{Department of Astronomy, University of Cape Town, Private Bag X3, Rondebosch, 7701, South Africa}

\author{M. J. I. Brown}
\affiliation{School of physics and Astronomy, Monash University, Clayton, Victoria 3800, Australia}

\author{Y. A. Gordon}
\affiliation{Department of Physics and Astronomy, University of Manitoba, Winnipeg, MB R3T 2N2, Canada}
\affiliation{Department of Physics, University of Wisconsin-Madison, 1150 University Ave, Madison, WI 53706, USA}

\author{S. Brough}
\affiliation{School of Physics, University of New South Wales, NSW 2052, Australia}

\author{A.M. Hopkins}
\affiliation{Australian Astronomical Optics, Macquarie University, 105 Delhi Rd, North Ryde, NSW 2113, Australia}

\author{B. W. Holwerda}
\affiliation{Department of Physics and Astronomy, 102 Natural Science Building, University of Louisville, Louisville KY 40292, USA}

\author{S. P.  Driver}
\affiliation{International Centre for Radio Astronomy Research (ICRAR), University of Western Australia, Crawley, WA 6009, Australia}

\author{E. M. Sadler}
\affiliation{Sydney Institute for Astronomy, School of Physics, University of Sydney, NSW 2006, Australia}
\affiliation{ATNF, CSIRO Astronomy and Space Science, PO Box 76, Epping, NSW 1710, Australia}
\affiliation{ARC Centre of Excellence for All Sky Astrophysics in 3 Dimensions (ASTRO 3D)}

%where ??1 = 9.52 × 1015 is the conversion factor from Mpc2 mJy to W Hz?1 , ?? is the radio spectral index13 , ?? is redshift, ?? ?? is the luminosity distance and ??1.4 is the measured 1.4 GHz flux density. We used the typical ?? = ?0.7 assumption (e.g. Kimball & Ivezi? 2008).

\begin{abstract}

The identification of AGN in large surveys has been hampered by seemingly discordant classifications arising from differing diagnostic methods, usually tracing distinct processes specific to a particular wavelength regime. However, as shown in Yao et al. (\citeyear{Yao2020}), the combination of optical emission line measurements and mid-infrared photometry can be used to optimise the discrimination capability between AGN and star formation activity. In this paper we test our new classification scheme by combining the existing GAMA-WISE data with high-quality MeerKAT radio continuum data covering 8 deg$^2$ of the GAMA G23 region. Using this sample of 1\,841 galaxies ($z<0.25$), we investigate the total infrared (derived from 12\micron) to radio luminosity ratio, \text{q$_{\text{(TIR)}}$}, and its relationship to optical-infrared AGN and star-forming (SF) classifications. We find that while \text{q$_{\text{(TIR)}}$} is efficient at detecting AGN activity in massive galaxies generally appearing quiescent in the infrared, it becomes less reliable for cases where the emission from star formation in the host galaxy is dominant. However, we find that the \text{q$_{\text{(TIR)}}$} can identify up to 70\,$\%$ more AGNs not discernible at optical and/or infrared wavelengths. The median \text{q$_{\text{(TIR)}}$} of our SF sample is 2.57\,$\pm$\,0.23 consistent with previous local universe estimates.

\end{abstract}


\section{Introduction}

Observations of galaxies at different wavelengths have shaped our understanding of their formation and evolution through time. However, the commonly derived parameters, such as stellar mass and star formation rate (SFR), rely on the assumption that the radiation received is exclusively generated by the stars within the galaxy. This assumption is true for pure star-forming (SF) galaxies, but not in the presence of an active galactic nucleus (AGN). 
AGNs are structures that radiate in the full electromagnetic spectrum, inducing additional flux to that emitted by stars. Although the most powerful (e.g., quasi-stellar objects) are easily identifiable, others with much weaker signatures can be hidden in the total emission from the host.

AGNs with a non-negligible jet kinetic power are referred to as radio loud (RLAGN) as opposed to the radio quiet AGN (RQAGN) showing lower power (e.g., Peacock et al. \citeyear{Peacock1986}; Padovani \citeyear{Padovani2011}; Bonzini et al. \citeyear{Bonzini2013}; Balokovic et al. \citeyear{Balokovic2014}).  The RLAGNs can be classified as radiatively efficient  (RE) or inefficient (RI)  based on their jet power, accretion rate and black hole mass, etc.  BL Lac objects are  RI objects in which the jets are oriented close to the line of site of the observer, while the low excitation radio galaxies (LERG) are seen at intermediate or large angle to the line of site. A RE system with a jet close to the line of site is  classified as a core-dominated, flat spectrum or OVV (optically violent variable) quasar; at intermediate and large angle, we have broad-line radio galaxies and narrow-line radio galaxies, respectively (see Table 1 in the review by Hardcastle and Croston \citeyear{Hardcastle2020} for more details). 

There are AGNs  where the accretion disk is hidden either by the dust/gas in the torus or  dust obscured star-forming regions within the host galaxy (known as obscured AGNs) making it invisible to the observer depending on the waveband used. Extensive work over the years have devised identification methods for obscured AGNs based on the following parameters:   X-ray hardness, the nuclear extinction from spectral analysis, the width of permitted emission lines, etc. (see the review by Hickox and Alexander \citeyear{Hickox2018}).

The main source of radio emission in galaxies is synchrotron radiation, which is a  non-thermal radiation caused by the acceleration of charged particles (mostly electrons) in the presence of a magnetic field. It is generated in star-forming (SF) regions by supernova remnants. The continuum radio emission is believed to be due to synchrotron emission by the relativistic electrons that rotate in the magnetic field carried by the jets in RLAGNs.  The physical process that generate radio emission in RQAGNs is still not well understood.  While some authors (Ulvestad et al. \citeyear{Ulvestad2005}) believe that synchrotron emission is due to less powerful jet in RQAGNs, others see shocks occurring in the accretion flow as the probable cause (see Ishibashi   $\&$ Courvoisier \citeyear{Ishibashi2011}).  The dynamics of accreting matter close to the central black hole likely  induces dissipative processes in the accretion flow, and shocks are naturally generated in such environments (Blandford \citeyear{Blandford1994}; Courvoisier $\&$ Turler \citeyear{Courvoisier2005}; Ishibashi $\&$ Courvoisier \citeyear{Ishibashi2009a}). Radio emission in RQAGNs is the main source of contamination that  causes  the departure from a linear far-IR/radio correlation (FIRRC).

Inside galaxies the optical and UV emissions from young and massive stars (M $>$ 8 M$_{\odot}$) are absorbed by their surrounding dust and reemitted in the infrared. On the other hand, the synchrotron radiation occurs within the supernova remnants left by these same massive and short-life stars ($\sim$\,Myrs vs Gyrs for solar-mass stars) when they die (Voelk  \citeyear{Voelk1989}; Lacki $\&$ Thompson  \citeyear{Lacki2010}). This cycle is used to explain the tight FIRRC  seen  for SF galaxies (de Jong et al. \citeyear{deJong1985};  Appleton et al. \citeyear{Appleton2004}; Ivision et al. \citeyear{Ivison2010}; Magnelli et al.  \citeyear{Magnelli2010}; Thomson et al.  \citeyear{Thomson2014}). 
This strong relationship between the radio and infrared continuum has been very useful in many studies, such as estimating the SFR in dusty nuclear starburst galaxies (Condon \citeyear{Condon1992}). In this way, the infrared/radio luminosity ratio q$_{\text{TIR}}$, can show excess either in FIR or radio luminosities, which is used to separate whether star formation or AGN activity dominates the emission. While the bulk of radio emission is synchrotron, there is some contribution from thermal free-free emission from \ion{H}{ii} regions (Condon \citeyear{Condon1992}).

At optical wavelengths, emission lines from the accretion disk are detected and specific combinations of these lines (Baldwin, Phillips and Terlevich \citeyear{Baldwin1981} known as BPT diagrams) have proven to be very useful for the separation  of AGNs and SF galaxies 
(see Baldwin, phillips and Terlevich \citeyear{Baldwin1981};  Veilleux et al. \citeyear{Veilleux1987}; Kewley et al. \citeyear{Kewley2001}; Kauffmann et al. \citeyear{Kauffmann2003}; Kewley et al. \citeyear{Kewley2006}; Yao et al. \citeyear{Yao2020}).  It does demand the presence of emission lines,  which may typically be faint to observe for most galaxies (e.g, the H-$\beta$ recombination line). 

The UV and optical emissions are absorbed in the dusty environment of the torus and re-emitted in the infrared.  Colors in this wavelength regime appear to be particularly efficient in discriminating between the different categories of galaxies (Jarrett et al. \citeyear{Jarrett2011}; Stern et al. \citeyear{Stern2012}; Jarrett et al. \citeyear{Jarrett2017}). The Wide-field Infrared Survey Explorer ($\textit{WISE}$,  Wright et al. \citeyear{Wright2010}) is an all-sky infrared survey in four bands namely W1(3.4\,$\mu$m), W2(4.6\,$\mu$m), W3(12\,$\mu$m), and W4(23\,$\mu$m).  W1 and W2 are both sensitive to the continuum emission from evolved stars, and the W2 band is additionally sensitive to hot dust; hence, this makes the 3.4\,$\mu$m - 4.6\,$\mu$m color an excellent diagnostic to identify galaxies dominated globally by AGN emission (see e.g. Jarrett et al. \citeyear{Jarrett2011}; Stern et al. \citeyear{Stern2012}; Assef et al. \citeyear{Assef2013}; Yao et al. \citeyear{Yao2020}).

However, all the AGN identification methods (see Padovani et al.  \citeyear{Padovani2017}) including the optical and infrared above-mentioned have a limited classification capability associated with the AGN's signatures and attributes probed. For example, a separation between AGNs and SF galaxies becomes very difficult when the AGN's host galaxy dominates the emission from the dusty torus (frequently the case for Seyfert-type galaxies). Similarly, optical lines become invisible when the AGN is completely obscured by dust, hence rendering the optical classification method ineffective. Studies have often found some X-ray AGNs among  SF  galaxies of the optical BPT classification (Trouille et al. \citeyear{Trouille2011}; Castell{\'o}-Mor et al. \citeyear{Castello2012}; Pons et al. \citeyear{Pons2016}; Agostino et al. \citeyear{Agostino2019}). Likewise, misclassifications are seen between optical BPT and infrared  $\textit{WISE}$ color-color (Ching et al. \citeyear{Ching2017}; Leahy et al. \citeyear{Leahy2019}; Yao et al. \citeyear{Yao2020}). Huang et al. (\citeyear{Huang2017}), using 18-band SED fitting in the mid-infrared, recover 20$\%$ more X-ray detected AGNs than using a mid-infrared color selection criteria. In contrast, Thorne et al. (\citeyear{Thorne2022}) employ far-UV to far-IR SED-fitting using Prospect, and find a significant AGN component in 91\% of galaxies selected using narrow emission line ratios and the presence of broad emission lines. They also find good agreement with mid-infrared color selections. 
On the other hand Hickox et al. (\citeyear{Hickox2009}; see their Figure 8)  found a relatively small overlap of about   30$\%$-50$\%$ between the X-ray and IR-selected AGNs in the AGES sample at 0.25 $<$ z $<$ 0.8 and the radio AGNs are generally not selected in the other wavebands. Regarding all the preceding issues one can see that no single waveband can  find all AGNs, but we rather need a multi-waveband approach.

\begin{figure*}[!th]
\centering
\includegraphics[width= 17cm]{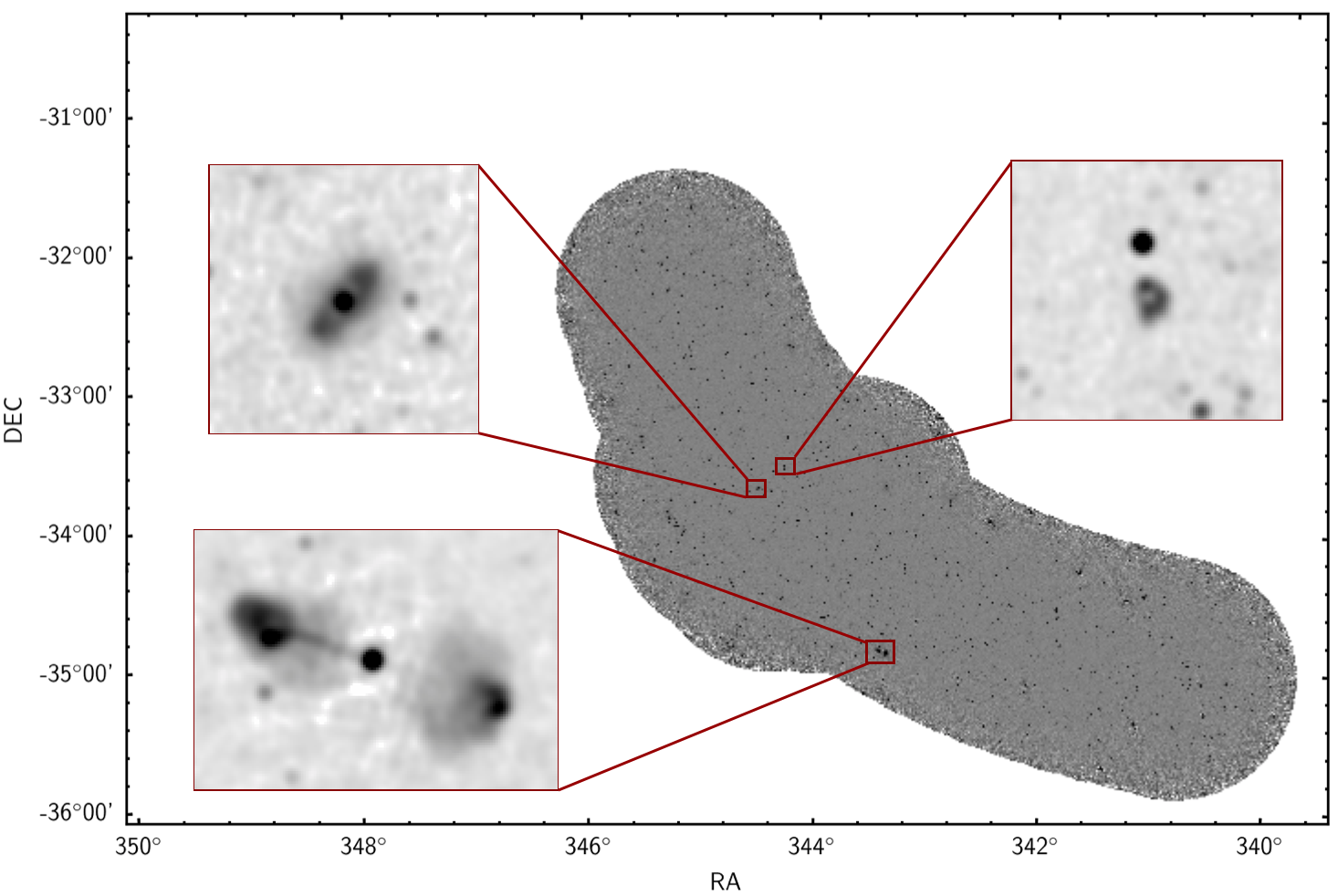}
 \caption{The MeerHOGS survey pilot region located in the GAMA G23 region. The greyscale shows the continuum mosaic at 1.4 GHz, as observed by MeerKAT (64 dishes). The observations targeted a filamentary structure between 0.025 $<$ $\textit{z}$ $< $ 0.034 with a dense galaxy group at its centre. The insets are close-up images of some well-resolved galaxies from our sample. Upper left: 3\,$\times$\,3 arcmin image of disk galaxy IC5271, lower left is a  5\,$\times$\,3 arcmin image of the giant radio galaxy PKS 2250--351 and upper right is a  radio galaxy (3\,$\times$\,3 arcmin image) which captured our attention owing to its circular-ring radio morphology. More resolved radio galaxies including the ones highlighted in this figure are presented with details in Figures  \ref{fig4} and  \ref{fig6.2b}.}\label{fig1}
\end{figure*}

In our previous paper (Yao et al. \citeyear{Yao2020} here referred to as ``Paper I''), we used robustly determined $\textit{WISE}$ photometry in combination with optical data from the Galaxy and Mass Assembly (GAMA, Driver et al. \citeyear{Driver2011}) to create a new ``hybrid" infrared-optical diagram --
combinations of H$\alpha$, [\ion{N}{II}] and WISE color --
which greatly increases the number of galaxies classifiable using the conventional BPT diagram (by a factor four).   
With the advent of large radio surveys on ASKAP and MeerKAT, and soon the SKA itself,
more studies will require quick and efficient ways to separate AGNs from SF populations without having to investigate the full energetic properties of these galaxies.    We are revisiting the classification of AGNs across the electromagnetic spectrum with the latest radio, optical and infrared  that we have available from the GAMA-G23 field, to investigate how well these classification methods can be exploited to improve upon the field. Our aim is to present a classification scheme (Section 5), which  shows the limitations as well as the power of combining these different wavelength regimes.

The structure of this paper is as follows: Section~\ref{sec2} describes the data used, including the radio continuum data reduction. In Section~\ref{sec3} we present the radio catalogue, while the radio properties of different optical-infrared classes of galaxies (derived in Paper I) are investigated in Section~\ref{sec4}. 
The discussion of our results is presented in Section~\ref{sec5}. Finally, Section~\ref{sec6} summarises our main findings.

The cosmology adopted throughout this Paper is H$_0$ = 70\,km s$^{-1}$ Mpc$^{-1}$, $\Omega_{M}$ = 0.3 and $\Omega_{\Lambda}$= 0.7. The conversions between the luminosity distance and the redshift use the analytic formalism of Wickramasinghe $\&$ Ukwatta (\citeyear{Wickramasinghe2010}) for flat, dark energy dominated Universe, assuming standard cosmological values noted above.
 All magnitudes are in the Vega system ($\textit{WISE}$ photometric calibration described in Jarrett et al. \citeyear{Jarrett2011}). Photometric colors are indicated using band names; e.g. W1 - W2 is the [3.4\,$\mu$m]  - [4.6\,$\mu$m]  color. Finally, for all four bands, the Vega magnitude to flux conversion factors are 309.68, 170.66, 29.05, 7.871\,Jy, respectively, for W1, W2, W3, and W4. Here we have adopted the W4 calibration from Brown et al. (\citeyear{Brown2014b}), in which the central wavelength is 22.8\,$\mu$m and the magnitude-to-flux conversion factor is 7.871\,Jy.

\section{Data} \label{sec2}

\subsection{Target Area}

The MeerKAT Habitat of Galaxies Survey (MeerHOGS) was executed as part of  the first MeerKAT call for proposals in early 2019 (Jozsa et al. 2022, Cluver et al. in prep.).  The observations covered a 10\,deg$^2$ area targeting a nearby ($\textit{z}$ = 0.03) filamentary structure of galaxies within the Galaxy and Mass Assembly (GAMA; Driver et al. \citeyear{Driver2011}) G23 field used for Paper I. It lies predominantly within the G23 region, except for a small portion at Declinations $<$ -35$^\circ$. The MeerHOGS area to be cross-matched with G23 is therefore reduced to $\sim$ 8\,deg$^2$.

Figure \ref{fig1} shows the survey's L-band continuum footprint and final mosaic, while the upper left, lower left, and upper right zoom insets are the galaxies IC5271, J2302-3718, and a ring-shaped galaxy that form part of our sample. These examples are taken from the resolved systems presented in Figure \ref{fig4} (fluxes provided in Table \ref{table2}), which will be discussed in the next section.

\subsection{Optical and Mid-infrared Data}

The primary data source used for this work is the optical and redshift data from the GAMA survey for the G23 region (Liske et al. \citeyear{Liske2015}, Baldry et al. \citeyear{Baldry2018}), combined with infrared photometry from the $\textit{WISE}$ survey (Wright et al. \citeyear{Wright2010}). This data is fully described in Paper I, however, we provide a brief overview here.

The optical line fluxes of the galaxies are extracted from the G23 GAMA II catalogue (GaussFitSimplev05 from within the SpecLineSFR Data Management Unit; Gordon et al. \citeyear{Gordon2017}). The spectra are fitted with the IDL code mpfitfun (Markwardt \citeyear{Markwardt2009}), which uses Levenberg-Marquardt non-linear least squares minimisation to identify the best-fitting parameters for the model, given the data and its associated uncertainties.

The mid-infrared photometry for the GAMA G23 region is constructed using the $\textit{ALLWISE}$ imaging frames which have been reprocessed to enhance the spatial resolution. A customised pipeline is used to carefully characterise and extract the photometry of well-resolved galaxies, which is then combined with data from the \textit{ALLWISE} catalogue for the remainder of the sample (see Cluver et al. \citeyear{Cluver2014}, Yao et al.  \citeyear{Yao2020} for details). To account for the redshifted emission across the optical and infrared bands, we applied a rest-wavelength correction  (i.e., ``k-correction") to the magnitudes
 based on spectral energy distribution (SED) fitting before the derivation of any physical value.

The optical spectra of the GAMA galaxies for which the W1 magnitude $>$ 15.5\,mag (0.2\,mJy) were visually inspected looking for potential blending (mostly broad [H$\alpha$] line blended with [\ion{N}{ii}]) of emission lines  in addition to taking into account the quality flag available, resulting in 9\,809 galaxies with high-quality photometry (the final sample or ``study sample"). These galaxies should constitute one of the best quality infrared-optical samples currently available. In this study we extend the power of this sample by combining it with the radio continuum data presented in Section 2.4.

\subsection{The Optical-Infrared Classifications of Galaxies} \label{New_diagram}

In Paper I, we carried out an extensive study of both the optical Baldwin, Phillips $\&$ Terlevich (\citeyear{Baldwin1981}) diagram  (BPT) and the mid-IR $\textit{WISE}$ color-color diagram, which led to the creation of a new AGN-SF classification method. It has the advantage of combining the W1$-$W2 color and the optical [\ion{N}{ii}]/H$\alpha$ line ratio, which proves to be a more efficient classification method than using $\textit{WISE}$ colors or optical line ratios alone. It allows the separation of galaxies into AGNs, pure SFs, and a ``mixed" class of composite galaxies believed to have AGN activity, but dominated by the star formation emission of the host galaxy (referred to as $\textit{Mixed}$ throughout this paper). A further classification is used for galaxies detected as AGN-dominated in the optical, but with significant star formation properties in the mid-IR: oAGN (mSF); see Paper I. In contrast, the AGNs in this classification scheme (Figure 14b in Yao et al. \citeyear{Yao2020}) are  a combination of the optical and infrared classifications; similarly the SF sample is better defined owing to the fact that it reflects both the BPT and the \textit{WISE} color classifications.

Earlier work using the $\textit{WISE}$ color-color diagram by Jarrett et al. (\citeyear{Jarrett2011}) (see also Stern et al. \citeyear{Stern2012}) derived color limits for AGN-dominated galaxies. These limits tend to be conservative as they were detecting mostly powerful infrared AGNs such as QSOs, obscured AGNs, and ULIRGs, yet missing the majority of low power AGNs. The AGNs from this class will be referred to as the ``Jarrett et al. (\citeyear{Jarrett2011}) classification".  Their properties will be compared to that of the AGNs in the new classification diagram, which has a much broader range of power. A summary table, which combines the different classification methods is presented in the discussion  (see Table \ref{summary_table}).

\subsection{Radio Continuum Data}

\subsubsection{Observations and Data Reduction}

The MeerHOGS project is one of the first radio continuum surveys using MeerKAT. The total observation time was 16.5\,hours (with 13.2\,h spent on source) over $\sim$ 10\,deg$^2$  at 1.4\,GHz from the 17th to the 31st of May 2019, with  25 pointings  in total (20 on the filament and an extra five in the dense galaxy group area).
The total observation time for each pointing was $\sim$ 30\,min. During each  observation  epoch, J2302-3718 was observed (2\,min) for the purpose of gain calibration. PKS 1934-63 was also observed (three times 10\,min) for bandpass and flux calibrations (see the observation details in Table \ref{MeerHOGS_observation}). The available frequencies range  from 900 - 1670\,MHz, but only a reduced interval from 1319.8 to 1517.1\,MHz (limited to redshifts $<$ 0.1) was used during the data reduction to minimise  the data volume and mitigate the RFI contamination.\\

\begin{table}[!h]
 \caption{Observation details of the MeerHOGS Survey} \label{MeerHOGS_observation}
\centering
\small{
\begin{tabular}{|l|c|c|c|}
\hline  
Date &    Number of antennas used       &Duration\\ \hline
17-03-2019 &  58&4.5 h\\\hline  
24-03-2019 &  58&4 h\\ \hline  
26-03-2019 &  64&4 h\\ \hline  
31-03-2019 &  58&4 h\\ \hline              
\end{tabular}
}
\end{table}

The data were reduced with the CARACal pipeline (J\'{o}zsa et al. \citeyear{Jozsa2020}). CARACal uses the Stimela Python framework, which can combine several versions of data reduction software in a system of  containers (see Chapter 4 of Makhathini, S. \citeyear{Makhathini2018} and references therein). The reduction process required 36 cores with a total memory of 300\,GB.

AOFlagger (Offringa \citeyear{Offringa2010}) was run to flag for shadowing and RFI. PKS 1934-63 (primary calibrator) was first used for the bandpass and gain calibration. Then, the result was transferred to the second calibrator for combined gain and flux calibration. The calibration tables obtained were finally applied to the observation and the continuum imaging was done using WSClean (Offringa et al. \citeyear{Offringa2014}). During the imaging process, the source-finding software SoFiA (Serra et al. \citeyear{Serra2015})  was used to generate a CLEAN mask using a threshold of $\sim$ 4\,$\sigma$ and following self-calibration with CUBICAL (Kenyon et al. \citeyear{Kenyon2018}). The final image resulted in a  synthesised beam of  13$.^{\prime\prime}$5$\times$13$.^{\prime\prime}$5 and an rms noise,   $\sigma_{\textrm rms}$, that ranges from 11 to 16$\mu$\,Jy.

\subsubsection{Source Extraction using $\textit{ProFound}$}   \label{section5b}

The source finding and extraction was performed using $\textit{ProFound}$ (Robotham et al.\,\citeyear{Robotham2018}). A threshold of  3\,$\sigma$, resulted in approximately  18\,000 sources (a source density of $\sim$ 1\,800 galaxies/deg$^{2}$). 
Our choice of $\textit{ProFound}$ was motivated by the fact that 
comparing $\textit{ProFound}$ to AEGEAN (Hancock et al. \citeyear{Hancock2012}, \citeyear{Hancock2018})  and pyBDSF (Mohan  \citeyear{Mohan2015}), Hale at al. (\citeyear{Hale2019}) found that $\textit{ProFound}$ has the ability to capture the complicated shape of the galaxies best. 

\begin{figure}
\centering
\includegraphics[width= 8.5cm] {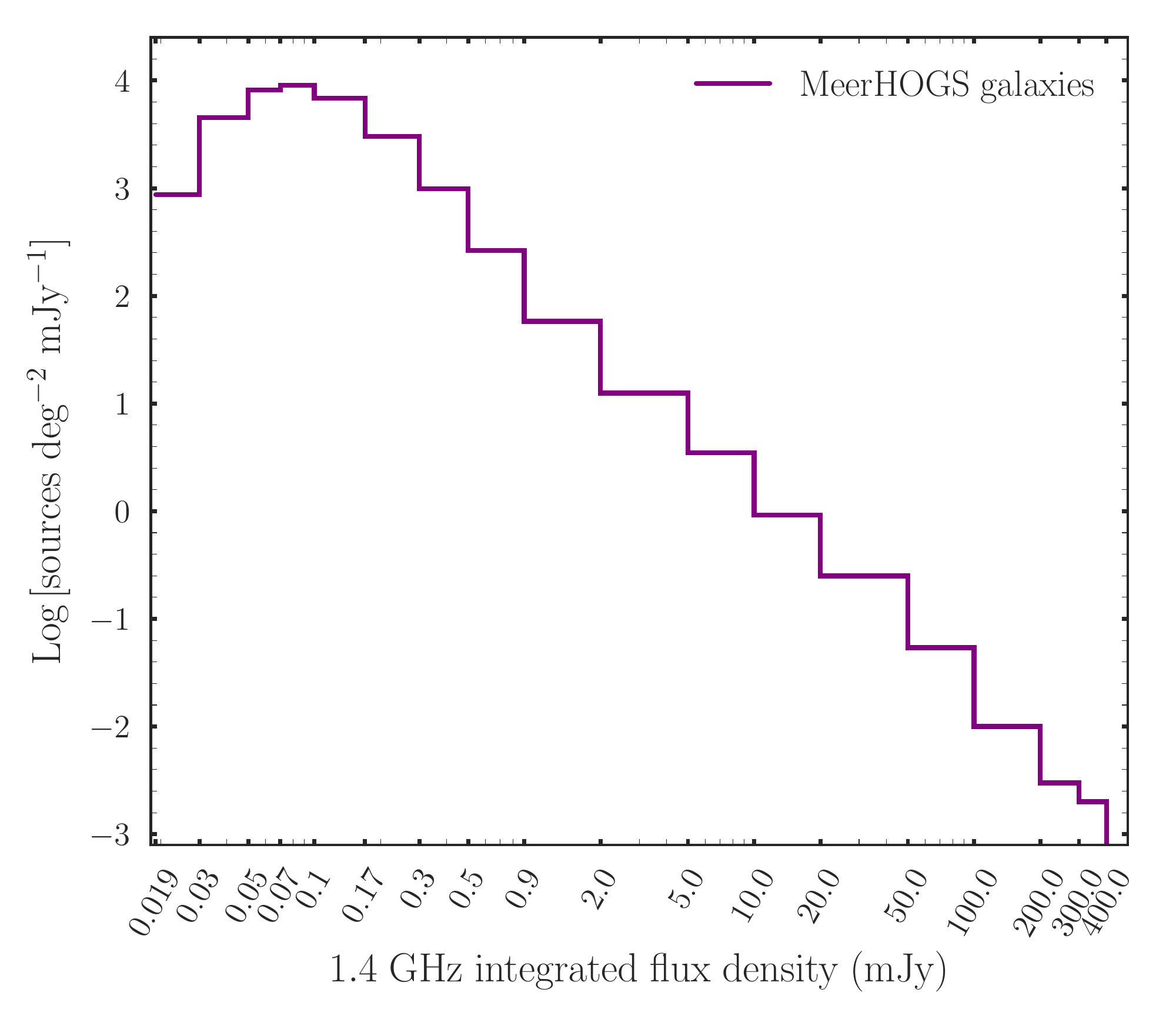}
\caption{Distribution of integrated 1.4\,GHz flux densities of all sources. The number of sources peaks around a flux density $\sim$ 0.1\,mJy. The two galaxies at the brightest end (0.3--0.4\,Jy) are respectively  MH5 and MH19 presented in Figure \ref{fig4} and Table \ref{table2}.} \label{fig2}
\end{figure}

The distribution of the 1.4 GHz integrated flux densities for all the sources is shown in Figure \ref{fig2}. There is a rise in the number of detections starting from flux density $\sim$ 0.03 mJy to 0.08 mJy, reaching $\sim$ 5000 galaxies at the peak.  It is followed by a constant (in the log) decrease in detections to a flux density of approximately 0.4 Jy.  The average rms noise is 13\,$\mu$Jy, which means we are reaching 6\,$\sigma$ at 0.08\,$\mu$Jy and about 3\,$\sigma$ for the faintest sources.

\begin{figure}
\centering
\includegraphics[width= 8.5cm] {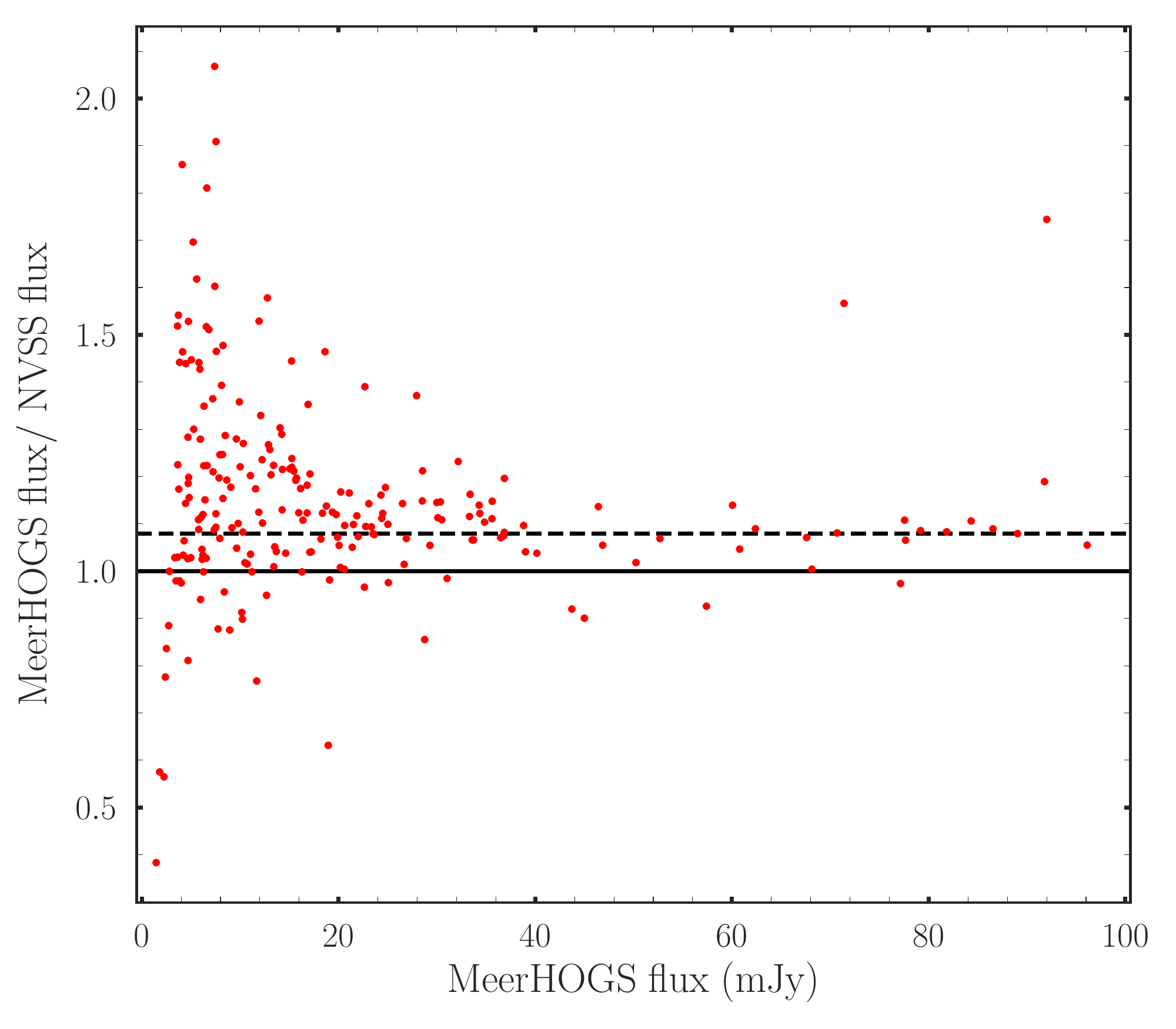}
\caption{MeerHOGS vs the ratio of MeerHOGS to NVSS flux density. The  figure shows a 8$\%$ systematic.  The MeerHOGS sources extracted using $\textit{ProFound}$ are on average $\sim$\,8$\%$ brighter than NVSS. This is likely related to the MeerHOGS calibration. Therefore, the median ratio of 0.92 was applied to correct the MeerHOGS  1.4\,GHz flux densities. The total number of galaxies shown is 213. } \label{fig3}
\end{figure}

\begin{table*}[!t]
 \caption{Radio flux densities of the resolved galaxies presented in Fig\,\ref{fig4}, including their component fluxes. The galaxies MH12 and MH14 were classified as three components although they are different from the conventional FRII.  The asterisks indicate galaxies with GAMA spectroscopic redshifts.} \label{table2}
\centering
%\hspace*{-0.2in}
\small{
\begin{tabular}{lccccccc}
\hline  \hline
Galaxy id.   &      RA        & Dec                &  $\mathrm{N. comp.}$               &  $\mathrm{East\;comp.}$             &  $\mathrm{central\;comp.}$  &  $\mathrm{West\;comp.}$ &  $\mathrm{Total\;flux}$\\
                   &            (deg)    &      (deg)              &                           &        (mJy)                       & (mJy)                    &      (mJy)               &(mJy)           \\   \hline
MH1  *    &   345.29553    &   -32.30024        &        1                      &   N/A              &   127.12             &   N/A                &   127.12   \\
MH2      &   344.22943    &   -32.70342        &        3                        &   11.35              &   3.64             &   15.18                &   30.18   \\
MH3  *    &   344.28134    &   -33.59379        &        1                      &   N/A              &   7.37             &   N/A                &   7.37   \\
MH4   *   &   344.50777    &   -33.74226        &        1                      &   N/A              &   20.92             &   N/A                &   20.92   \\
MH5   *   &   343.40012    &   -34.92519        &        3                      &   139.35              &   41.4             &   112.33                &   293.08   \\
MH6       &   345.29649    &   -33.93349        &        3                       &   70.96              &   3.87             &   95.99                &   170.83   \\
MH7      &   344.30609    &   -34.54677        &        3                        &   31.56              &   3.06             &   35.71                &   70.33   \\
MH8      &   345.21566    &   -34.02112        &        1                        &   N/A              &   30.71             &  N/A                &   30.71   \\
MH9     &   345.4142     &   -33.26196         &        1                         &   N/A              &   57.4             &   N/A                &   57.4   \\
MH10  &         344.81651 &     -32.2304         &       1                       &   N/A              &   4.32             &  N/A               &   4.32   \\
MH11  &         345.2202  &     -32.84014        &    3                          &   2.18              &   0.87             &   0.98                &   4.04   \\
MH12  &       345.53025  &    -33.36734         &   3                           &   6.39              &   10.55             &   5.79                &   22.73   \\
MH13  &       345.20294  &    -33.14328         &   2                           &   1.45              &   N/A            &   6.66                &   8.12   \\
MH14  &        344.57336  &    -32.97658        &   3                           &   2.98              &   18.38             &   2.1                &   23.46   \\
MH15  &        343.64585  &    -33.09938        &   2                           &   7.37              &   N/A             &   3.95                &   11.32   \\
MH16 &         342.9438  &     -33.33359        &   3                            &   6.96              &   2.37             &   16.66                &   26.0   \\
MH17 &         345.17558 &     -34.08114       &   2                            &   28.06              &   N/A             &   13.84                &   41.9   \\
MH18 &         343.02919 &     -34.206           &   3                            &   9.57              &   N/A            &   5.22                &   14.78   \\
MH19 &         341.66499 &     -34.32315       &   2                            &   217.29              &   N/A             &   88.42                &   305.72   \\
MH20 *&        345.09855 &     -31.69654       &   3                           &   1.5              &   3.62             &   3.87                &   8.99     \\  \hline

\end{tabular}
}
\end{table*}

In Figure \ref{fig3} we compare the flux measurements from the MeerHOGS  with that of the NRAO VLA Sky Survey (NVSS; Condon et al. \citeyear{Condon1998}) catalog data. \color{black} 
The measurements from MeerHOGS are systematically  brighter than that of the NVSS by 8$\%$. 
  A reduced sample of twelve galaxies, presented in Healy et al. (\citeyear{Healy2021}; Figure 2) using the fractional difference between MeerKAT and NVSS fluxes, shows  a similar trend (except for the systematic  offset in flux). However,  they used  an in-house NRAO software package called VSAD (see Section 5.2 of Condon et al. \citeyear{Condon1998}) for their source extraction instead of $\textit{ProFound}$. The scatter seen in our Figure \ref{fig3} could be related to  the difference in sensitivity between the two surveys.  But, $\textit{ProFound}$, whose model fit follows the distribution of the emission for each galaxy, is likely collecting slightly more flux as opposed to the extraction methods used for the  NVSS, thereby contributing to the observed 8$\%$ systematic offset.  It is also, however, likely that the MeerKAT flux calibration is a contributing factor. To ensure consistency with previous studies, we apply a correction factor of 0.92 to bring our data in line with that of NVSS, allowing direct comparisons between our results and the findings of others.

\section{The MeerKAT Radio Continuum Catalogue} \label{sec3}

\begin{figure*}[!ht]
\centering
\includegraphics[width= 3.9cm] {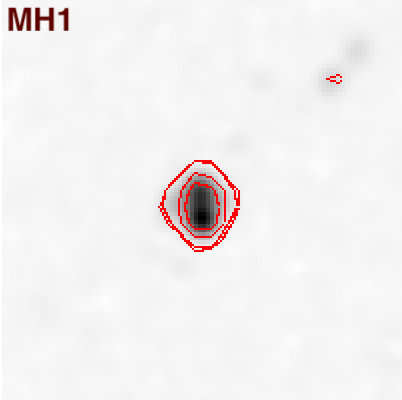} 
\includegraphics[width= 3.9cm] {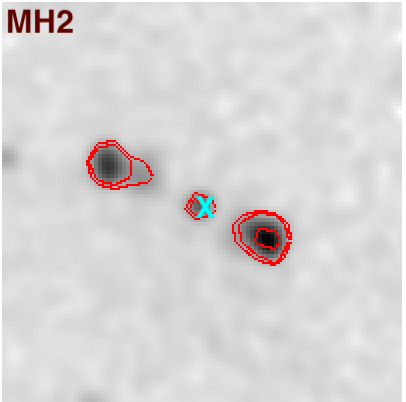} 
\includegraphics[width= 3.9cm] {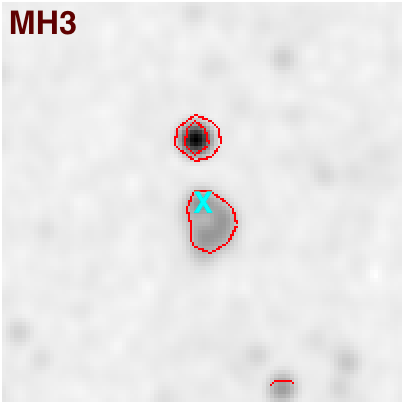}
\includegraphics[width= 3.9cm] {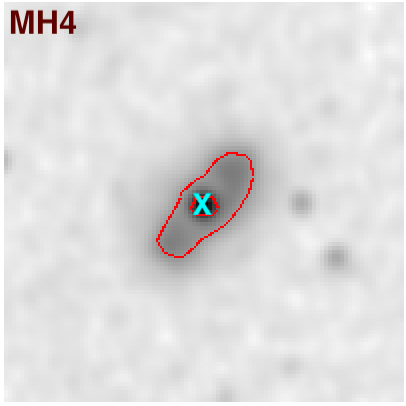}\\ 
\includegraphics[width= 3.9cm] {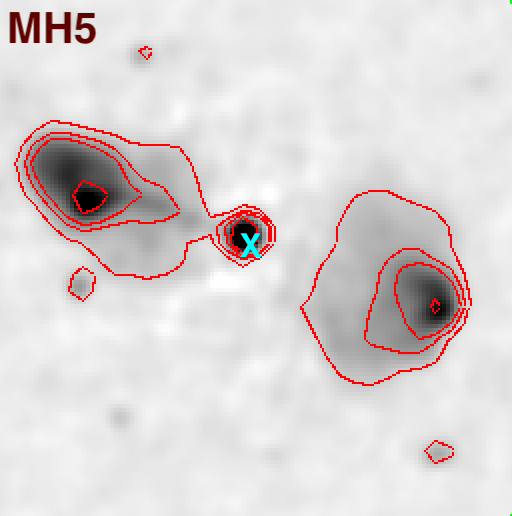}
\includegraphics[width= 3.9cm] {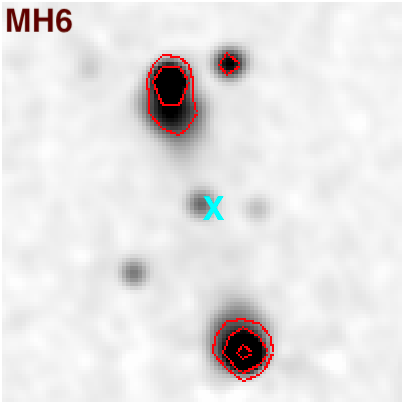} 
\includegraphics[width= 3.9cm] {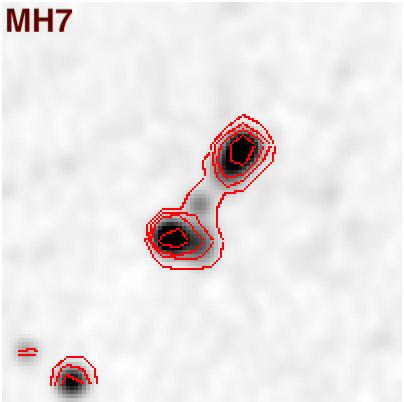}
\includegraphics[width= 3.9cm] {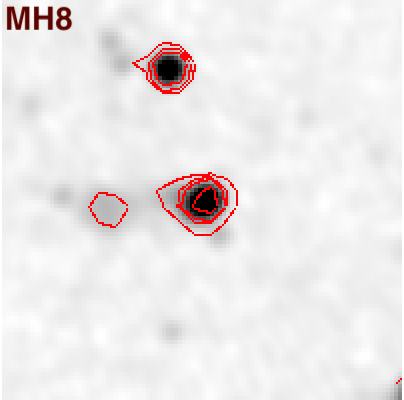}\\
\includegraphics[width= 3.9cm] {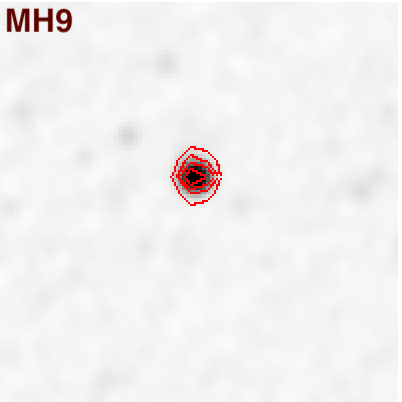}
\includegraphics[width= 3.9cm] {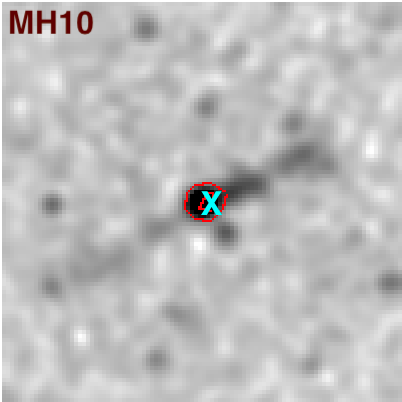}   
\includegraphics[width= 3.9cm] {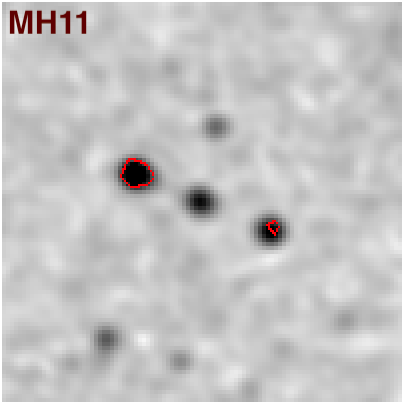} 
\includegraphics[width= 3.9cm] {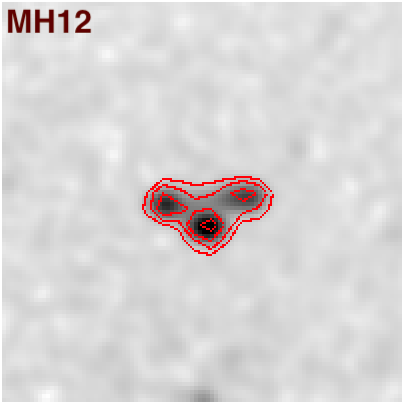}\\ 
\includegraphics[width= 3.9cm] {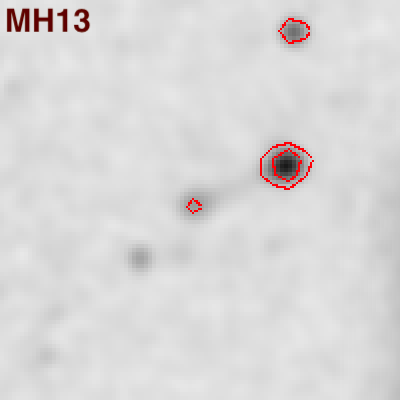} 
\includegraphics[width= 3.9cm] {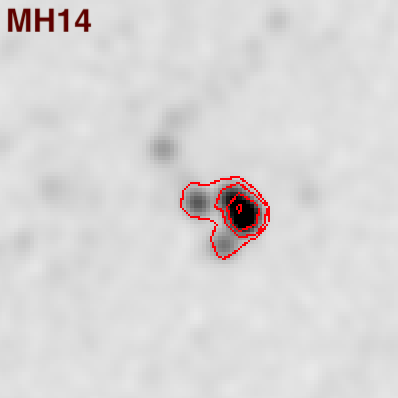}
\includegraphics[width= 3.9cm] {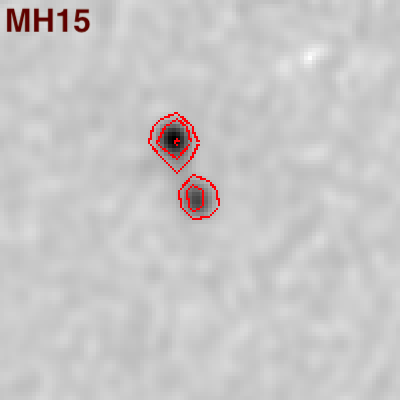}
\includegraphics[width= 3.9cm] {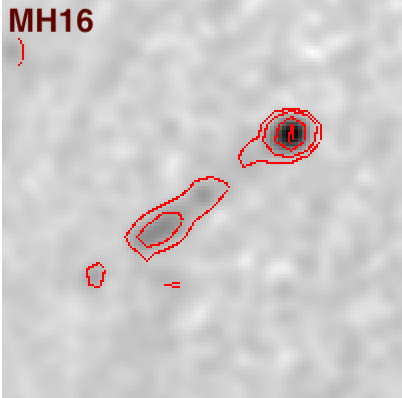}\\ 
\includegraphics[width= 3.9cm] {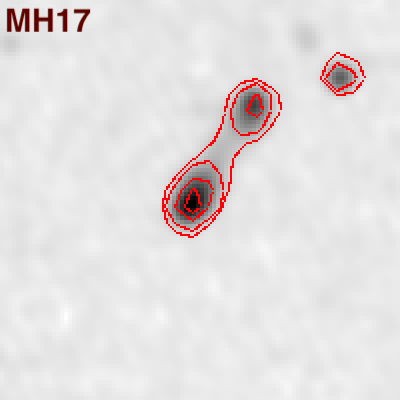} 
\includegraphics[width= 3.9cm] {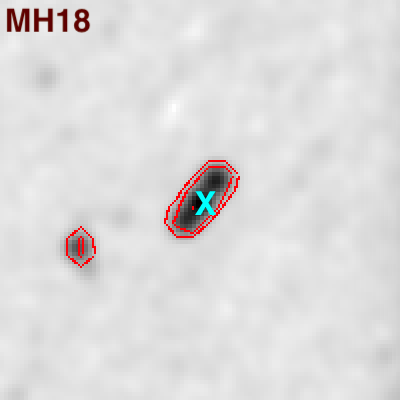}
\includegraphics[width= 3.9cm] {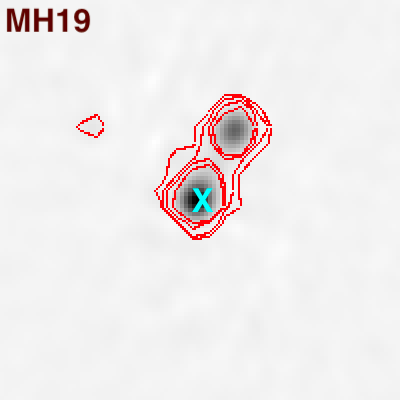}
\includegraphics[width= 3.9cm] {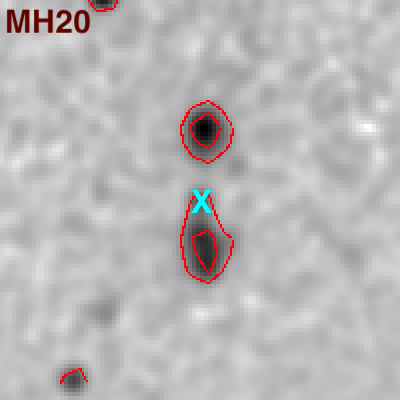}\\
\caption{A representative selection of resolved galaxies in the MeerHOGS region. Each MeerKAT radio continuum stamp is 5$\times$5$^\prime$ (except MH5, which is 7$\times$7$^\prime$) and the beam size is 13.5$^{\prime\prime}$. See Table\,\ref{table2} for measurements. MH5 is a known giant radio galaxy (GRG), which will be discussed in the text. The contours are 20\,$\mu$Jy, 50\,$\mu$Jy, 150\,$\mu$Jy and 300\,$\mu$Jy. The cyan crosses indicate the positions of the \textit{WISE}  galaxy catalogue counterparts. See Figure \ref{fig6.2b} for additional sources.}  \label{fig4}
\end{figure*}

\subsection{Resolved Sources}

We extracted $\sim$18\,000 radio sources from the MeerHOGS MeerKAT mosaic. The majority of detections are, however, point sources with relatively few ($\sim$ 170) resolved galaxies (i.e. sources with dimensions at least twice the circular beam size of 13$.^{\prime\prime}$5). A selection of objects taken from the radio resolved population is presented in Figure \ref{fig4}, with their fluxes given in Table \ref{table2}. This list of resolved objects is, however, based on visual inspection of the extracted apertures and continuum imaging; apart from the obvious cases of FRII sources such as MH2 or MH5 (see Figure \ref{fig4}), some of these may therefore be a chance superposition of sources  and hence not necessarily physically associated. 

FRII objects are luminous radio sources with hotspots in their lobes at the edge of the jets (see  Fanaroff $\&$ Riley  \citeyear {Fanaroff1974}). Sources with triple components like MH2, MH6, MH7 etc., \color{black} are likely FRIIs for which the central AGN is visible along with the lobes, as well as some complex systems with three radio components. FRII sources for which the central part is faint, or no longer visible, constitute the majority of the double components. Some exhibit one very bright lobe in comparison to the other; this is likely the result of an orientation effect boosting the jet pointing toward us at a low angle. This effect might be exacerbated when the angle is close to zero. Indeed, many single component sources and extremely bright galaxies in our radio catalogue (about 50 cases; e.g. MH8 and MH9) seem not to have a counterpart in $\textit{WISE}$, which means they are either emitting weakly in the IR or are distant galaxies whose directly pointing jets are being Doppler-boosted. They are intriguing galaxies that will need further attention with deeper spectroscopic data. 

In addition we find some galaxies with abnormal morphology, such as MH3, which shows a bent shape probably caused by an interaction of the jet with its surrounding environment, or possibly a galaxy merger.

\begin{table*}[!thb]
 \caption{Radio flux densities  of the components of MH5 (PKS 2250--351).  Some of the measurements  compiled by Seymour et al. (\citeyear{Seymour2020})  are presented along with our own measurements using MeerKAT (64 dishes).} \label{table3}
\centering
\small{
\begin{tabular}{lcccccc}
\hline  \hline
Telescope  &      Survey         & Frequency        &  East Lobe                    &        Core                  &  West Lobe  & Total \\
                   &                          &      (GHz)           &      (mJy)                       &        (mJy)                  & (mJy)                       &      (mJy) \\   \hline
ATCA          &   Green Time    &   9.5                   &        22.1 $\pm$ 1.2      &     66.2  $\pm$ 3.3     &  10.7$\pm$ 0.6                &  96.6 $\pm$ 3.4 \\
ATCA          &   Green Time    &  5.5                    &        38.7 $\pm$ 2.1      &     71.4 $\pm$ 3.6       &   23.6 $\pm$1.4               &   134 $\pm$ 49  \\
VLA            &  NVSS               &   1.4                   &        135 $\pm$ 5           &     50.0 $\pm$ 2.2        &    100.1 $\pm$ 3.8           & 285 $\pm$ 29\\
\textbf{MeerKAT}     &  \textbf{MeerHOGS}    &   \textbf{1.4}                   &    \textbf{139.38 $\pm$ 0.1}  &     \textbf{41.4 $\pm$ 0.1}       &    \textbf{112.33 $\pm$ 0.4}  &  \textbf{293.112 $\pm$ 0.6} \\ 
ASKAP       &    EMU              & 0.888                  &        193 $\pm$ 26        &       N/A                               &            145 $\pm$17    &  $>$ 338 \\
MOST        &   SUMSS           &   0.843               &       175.1  $\pm$ 5.8    &      64 $\pm$ 5.6           &      153.8 $\pm$ 7.5     &   393 $\pm$ 41\\
uGMRT      &   GLASS            &   0.675              &         237 $\pm$ 17        &   45 $\pm$ 3               &182 $\pm$ 12                 &  464 $\pm$ 29 \\ \hline                   
\end{tabular}
}
\end{table*}

The largest angular source in our continuum imaging is MH4, identified as IC5271. It is a well-resolved (118$^{\prime \prime}\times$70$^{\prime \prime}$) disk galaxy in the radio and similarly in $\textit{WISE}$ (a $\textit{WISE}$ 3-color image and other characteristics are presented in Appendix A, Figure \ref{fig6.14}). Such galaxies, resolved and large enough to do a pixel-by-pixel study, are rare, and MH4 is the only one in our entire sample.
 
It is followed by MH5, the famous PKS 2250--351, a radio giant galaxy for which flux measurements are available in several works (Brown et al. \citeyear{Brown1991};  Condon et al. \citeyear{Condon1998}; Intema et al.  \citeyear{Intema2009}; and most recently, using ASKAP, Seymour et al. \citeyear{Seymour2020}). We found a total flux of $\sim$ 293.112\,$\pm$\,0.6\,mJy, which is similar to that found by the NVSS  (285\,$\pm$\,29\,mJy).

\begin{figure}
\hspace*{0.4in}
\includegraphics[width= 7.1cm] {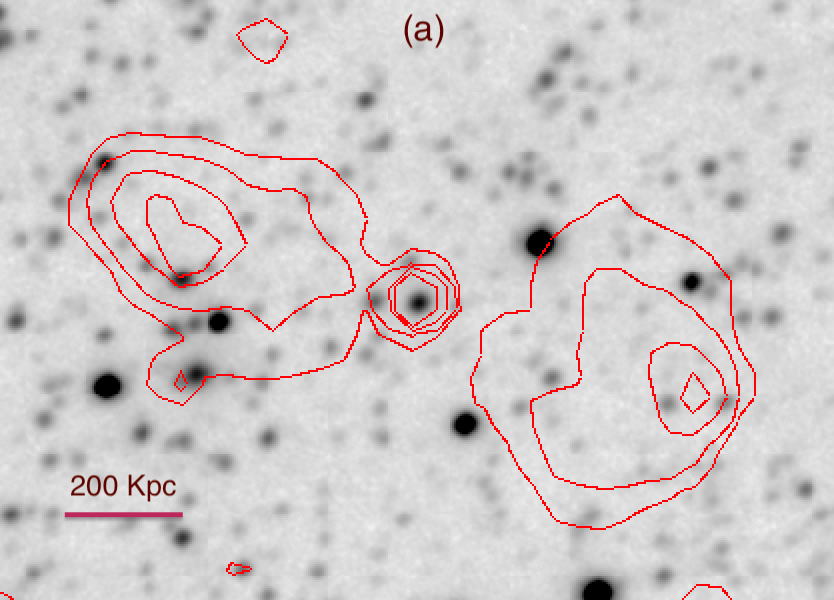}\\
\hspace*{-0.12in}
\includegraphics[width= 8.8cm] {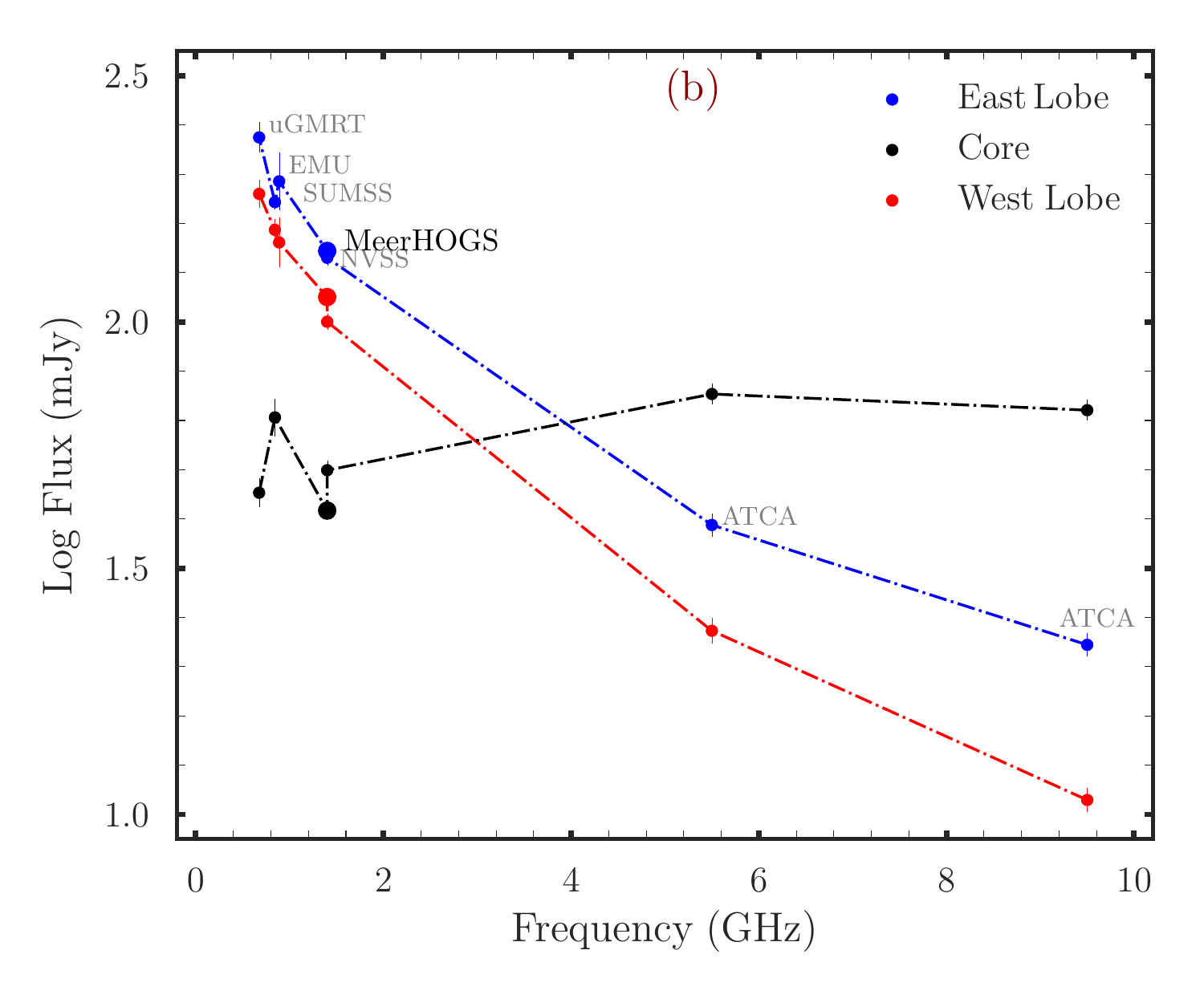}\\
\caption{The giant radio galaxy (GRG) PKS 2250--351. (a): greyscale W1\,(3.4\,$\mu$m) image of the galaxy  with 1.4 GHz image contours overlaid in red (upper panel). The contour levels are 20\,$\mu$Jy, 50\,$\mu$Jy, 150\,$\mu$Jy and 300\,$\mu$Jy, respectively. (b) shows the variation of flux in the different components  (East lobe, Core, West lobe) of  the GRG (MH5) as presented in Table \ref{table3}. The three components of the galaxy are measured in all the surveys except for EMU, where there is no measurement for the core. The larger point is used to indicate this MeerHOGS data to distinguish it from NVSS. The flux density in the core remains flat, while it decreases in the lobes with higher frequencies. It shows a steep synchrotron signature. Given the large size of MH5, we can postulate multiple shocks during the expansion of the jets as a potential cause of the steep spectrum as suggested by Gopal-Krishna $\&$ Wiita (\citeyear{Gopal1990}).  \label{fig5}} 
\end{figure}

For resolved FRII galaxies like MH5, a visual inspection of the segmentation map is needed to sum the flux contributions from the lobes and the central galaxy to determine the total flux of the galaxy (see an example of segmentation map in Figure 3 of Robotham et al. \citeyear{Robotham2018}).
The jets of MH5 extend across $\sim$350\arcsec (1.25\,Mpc) from end to end. We reproduce for completeness several of the relevant measurements compiled by Seymour et al. (\citeyear{Seymour2020}) in Table \ref{table3} and plotted in Figure \ref{fig5}.   Consistent with steep-spectrum synchrotron emission, we see a rapid decrease (low to high frequency) of the flux in the two lobes West (red) and East (blue) from $\sim$ 200\,mJy at 675 MHz to $\sim$ 10\,mJy at 9.5\,GHz, while the emission from the core remains roughly constant (from 45 to 66\,mJy). MH5 (PKS 2250--351) is also used as a comparative example of a powerful AGN in the analysis that follows.\\

\subsection{MeerHOGS Cross-matched with the GAMA-\textit{WISE} Sample} \label{MeerHOGS-Cross-matched}

\begin{figure}[!ht]
\centering
\includegraphics[width= 8.5cm] {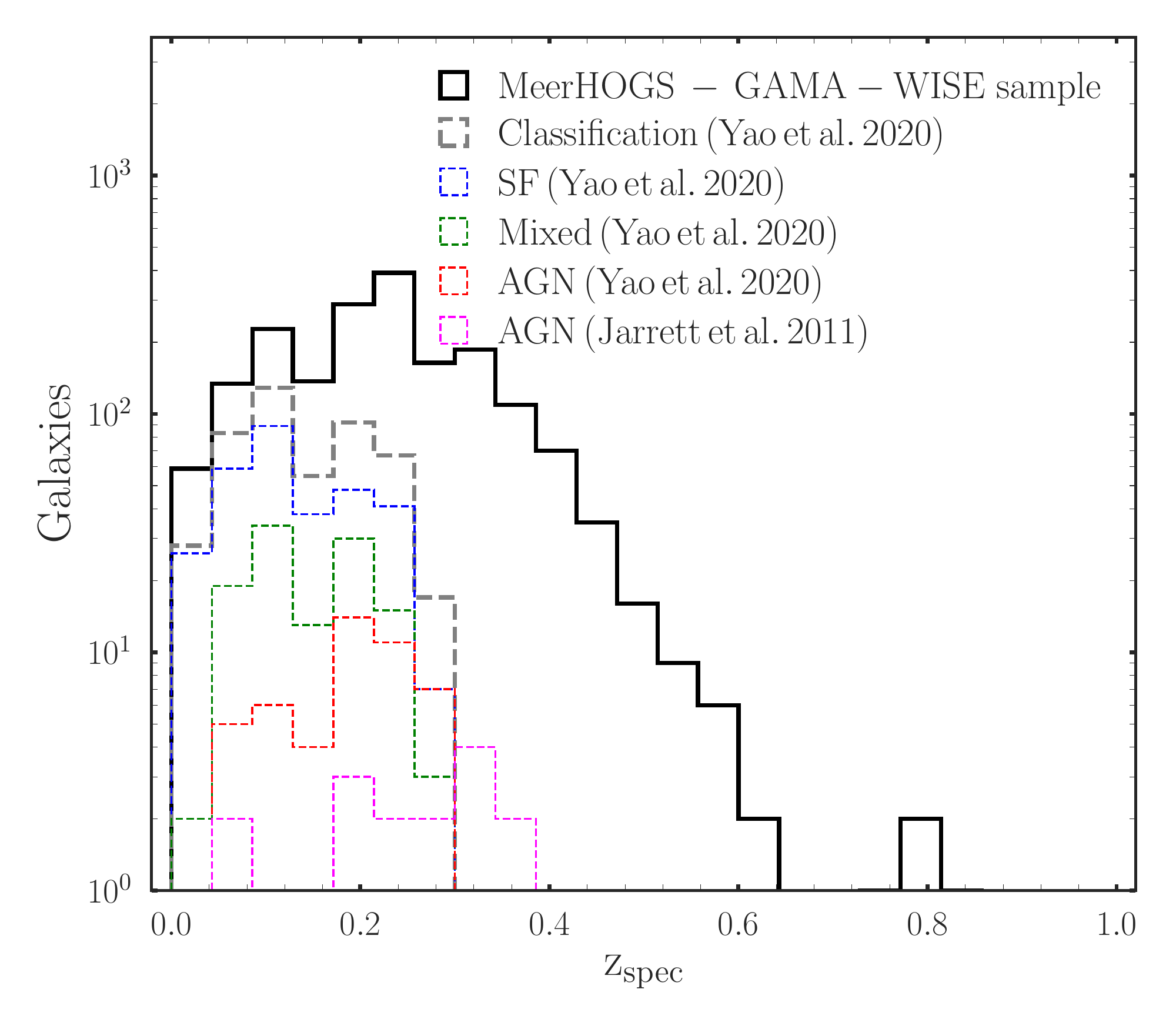}\\
\caption{The redshift distribution of the radio galaxies in MeerHOGS. The total Radio--Optical--IR sample is represented in black. We can see radio galaxies all the way to redshift $\textit{z}$ $\sim$ 0.6, but the bulk ($>$75$\%$) of the distribution is at $\textit{z}$ $<$ 0.3, which corresponds to the redshift limit applied to the GAMA-$\textit{WISE}$ study in Paper I. The distributions of the galaxies classified by our new classification diagnostic  (Yao et al. \citeyear{Yao2020}) are also presented, along with the AGNs classified using the $\textit{WISE}$ color method by Jarrett et al. (\citeyear{Jarrett2011}). All the classifications from  Yao et al. (\citeyear{Yao2020}) are based on the GAMA-$\textit{WISE}$ sample  limited to $\textit{z}$ $<$ 0.3.} \label{fig6}
\end{figure}

A cross-match of the radio data with the GAMA-\textit{WISE} sample using a 5$^{\prime\prime}$ search radius resulted in 1\,841 radio galaxies having a $\textit{WISE}$ counterpart and a redshift available in GAMA (here referred to as the ``MeerHOGS--GAMA--$\textit{WISE}$ sample" or total sample). 
We assess the reliability of our cross-match  by making use of the method described in Ching et al. (\citeyear{Ching2017}; see section 3.6), which is based on Monte Carlo simulations. The basic procedure is as follows: We first define the domaine of possible events and generate them randomly. Then, we  perform deterministic judgments of the system states based on the events and finally count the occurrence of a specific system state among total observations (see Matsuoka \citeyear{Matsuoka2013}). In the current study the reliability (R) is defined as the probability that a matched radio source is genuinely associated to its counterpart rather than a random source on the sky. R = (1 - \big<N$_{\text{rand}}^{\text{match}}$\big>/N$_{\text{true}}^{\text{match}}$), where N$_{\text{true}}^{\text{match}}$ is the number of matches from the true catalogue, and \big<N$_{\text{rand}}^{\text{match}}$\big> is the average number of matches using the random catalogues. Based on the preceding definition, we find  89$\%$ reliability using a  5$^{\prime \prime}$ cross-match radius between $\textit{WISE}$ and MeerHOGS positions. A 3$^{\prime \prime}$ cross-match radius leads to  a higher reliability of 93.5$\%$ (for a 1$^{\prime \prime}$ it is 97.6$\%$), but would result in the loss of  $\sim$30$\%$ of our current sample.  Repeating the analysis in this paper with 3$^{\prime \prime}$ cross-match radius has a negligible impact on our analysis and leaves our findings unchanged.  We additionally checked the possibility of having multiple $\textit{WISE}$ sources within 5$^{\prime \prime}$ of each MeerHOGS source. Only 45 cases (from a total of 1841 galaxies, $\sim$ 2$\%$) of multiple sources were found. Similarly only 2 cases ($\sim$0.1$\%$) of multiple MeerHOGS sources were found within 5$^{\prime \prime}$ around the position of $\textit{WISE}$ sources.
The positional offsets between MeerHOGS and $\textit{WISE}$ galaxies for 5$^{\prime \prime}$ cross-match radius are presented in the Appendix (Figure \ref{fig6.1b}), showing excellent agreement.

In Paper I we devised a new classification scheme that identifies AGNs by combining optical excitation ([\ion{N}{ii}] and H$\alpha$) and the W1 $-$ W2 color. The main criteria for this classification are [\ion{N}{ii}]  and H$\alpha$ lines in emission with S/N $>$ 3, S/N (W1) $>$ 5, and S/N (W2)  $>$  5, respectively.  478 (478 from the total sample 1841 galaxies) galaxies from the MeerHOGS--GAMA--$\textit{WISE}$ sample satisfy these conditions. 

The redshift distribution of the parent MeerHOGS--GAMA--$\textit{WISE}$ sample is presented in Figure \ref{fig6}. The total sample (black) is distributed from $\textit{z}$ $\sim$ 0 up to $\textit{z}$ $\sim$ 0.8, but no more than 2\,$\%$ of the galaxies are found beyond $\textit{z}$ = 0.5 (the nominal redshift limit of GAMA) and 76\,$\%$ have $\textit{z}$ $<$ 0.3 (corresponding to the median redshift of the GAMA survey) that was used to derive the combined classification scheme. The SF and $\textit{Mixed}$ galaxies using the Yao et al. (\citeyear{Yao2020}) classification have a flat distribution between $0 \le z \le 0.3$, but the AGNs appear to be more concentrated at $\textit{z}$ $\sim$ 0.25, consistent with being rare and more readily detected in larger volumes (i.e. at greater distance). The AGNs classified by the method of Jarrett et al. (\citeyear{Jarrett2011}), which are generally very luminous in the mid-IR are found to $z<0.4$. The star/galaxy separation algorithm used by the GAMA survey biases against QSOs, and they are hence rare in the GAMA catalogue (e.g. Baldry et al. \citeyear{Baldry2010}).

\section{Optical--IR--Radio Study of AGN vs SF} \label{sec4}

In this section, the MeerHOGS--GAMA--$\textit{WISE}$ sample is used to study the properties of  AGNs and SF galaxies. Parameters such as stellar mass and \text{SFR$_{ {12\mu}\text{m}}$} are already available from our $\textit{WISE}$-GAMA catalogue (see Paper I for details).  Our aim here is to combine radio, optical, and infrared information to better classify galaxies whose hosts may contain AGN.

\subsection{The Infrared-to-Radio Luminosity Ratio: q$_{\text{(TIR)}}$}

Several studies in the literature have established a tight correlation between the far-infrared (FIR) and the non-thermal radio flux density (at 1.4\,GHz) for star-forming galaxies (Van der Kruit \citeyear{Van1971}; Helou et al. \citeyear{Helou1985}; Condon et al. \citeyear{Condon1992}), which appears to hold also at mid-IR wavelengths (Appleton et al. \citeyear{Appleton2004};  Huynh et al. \citeyear{Huynh2010}).
The q ratio, defined as the infrared divided by the radio luminosity, can be used to probe the nature of the dominant process heating the interstellar medium, thus helping to classify galaxies as starbursts or AGNs based on the excess IR or radio emission, respectively.

In this study we derive the total infrared (TIR)/radio luminosity ratio, q$_{\text{(TIR})}$, using  12\,$\mu${m} as a proxy for TIR -- indicated here using (TIR). As shown by Cluver et al. \citeyear{Cluver2017}) the  rest-frame luminosity from the W3 band closely tracks TIR luminosity, even in the presence of mid-infrared AGN, with a tight correlation ($\sim$ 15\,$\%$ scatter) between the two parameters given by Equation \ref{eq6.0}:\\

\begin{multline} \label{eq6.0}
\text{log$_{10}$}\,\text{L$_{\text{(TIR)}}$}(\text{L$_{\odot}$})= (0.889 \pm 0.018) \text{log$_{10}$}\,\text{L$_{ {12\mu}\text{m}}$}\,(\text{L$_{\odot}$})\\
 + (2.21 \pm 0.15)\\
\end{multline}

The 1.4\,GHz radio flux densities were converted to rest-frame 1.4\,GHz effective luminosities  in Equation \ref{eq6.21} assuming a radio spectral index of $\alpha$ = - 0.7 (Ibar et al.  \citeyear{Ibar2010}).

\begin{equation}  \label{eq6.21}
\text{L$_{\text{1.4GHz}}$} =   4{\pi}d_{lum}^2 \frac{\text{S$_{\text{1.4GHz}}$}}{\text{(1+z)}^{1+\alpha}}
\end{equation}

The q$_{\text{TIR}}$ ratio is defined in Equation \ref{eq6.2} (from Helou et al. \citeyear{Helou1985}):
 
 \begin{equation}  \label{eq6.2}
\text{q$_{\text{TIR}}$} = \text{log$_{10}$}\,\bigg({\frac{{\text{L$_{\text{TIR}}$}}}{3.75\times10^{12}\text{Hz}}}\bigg) -  \text{log$_{10}$}\,\bigg(  {\frac{\text{L$_{\text{1.4GHz}}$}}{\text{W Hz}} }\bigg)
\end{equation}

An excess of \text{q$_{\text{TIR}}$} is an unambiguous signature of  radio-loud AGNs, whereas radio-quiet AGNs follow the same radio-FIR relation as SF galaxies. \text{q$_{\text{TIR}}$} is therefore an incomplete, albeit important, diagnostic.

\begin{figure}
\includegraphics[width= 8.5cm] {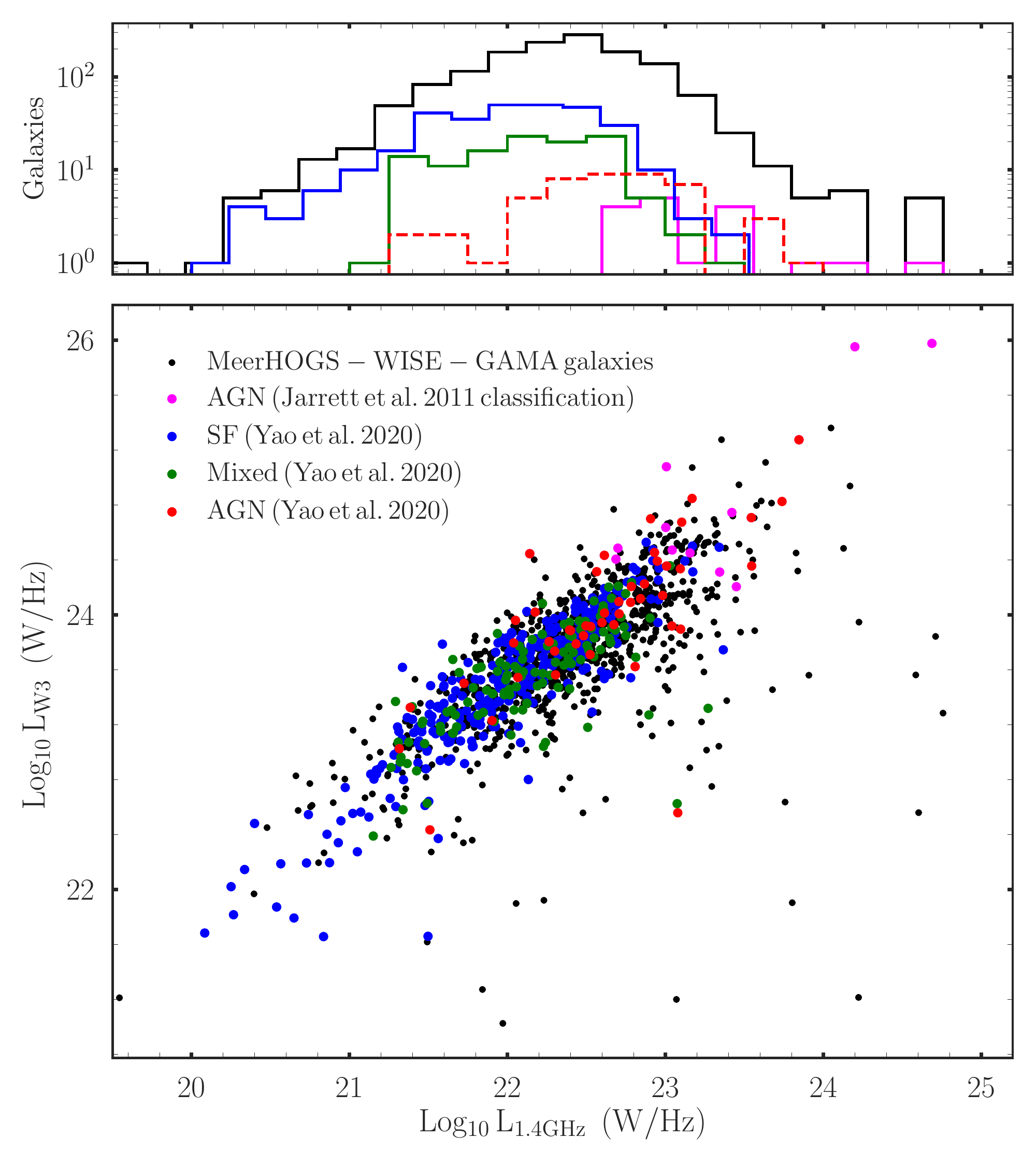} 
\caption{Variation of the luminosity W3 band versus the radio luminosity (L$_{\text{1.4GHz}}$). There is a correlation between the two luminosities for our sample of galaxies.  The total sample, the SFs, the Mixed, and the AGNs are represented in black, blue, green, magenta, and red, respectively (see the corresponding histograms on the top panel.} \label{fig7}
\end{figure}

 In Figure \ref{fig7} we show the W3 luminosity vs radio luminosity for the total sample. It includes all galaxy types,  but a clear correlation between the two luminosities is evident.
The high  luminosities seen in the infrared and the radio for galaxies classified as AGNs (red and magenta)  are probably  dominated by the central AGN rather than being associated with evolving stars.

\begin{figure*}
\centering
\includegraphics[width= 8.5cm] {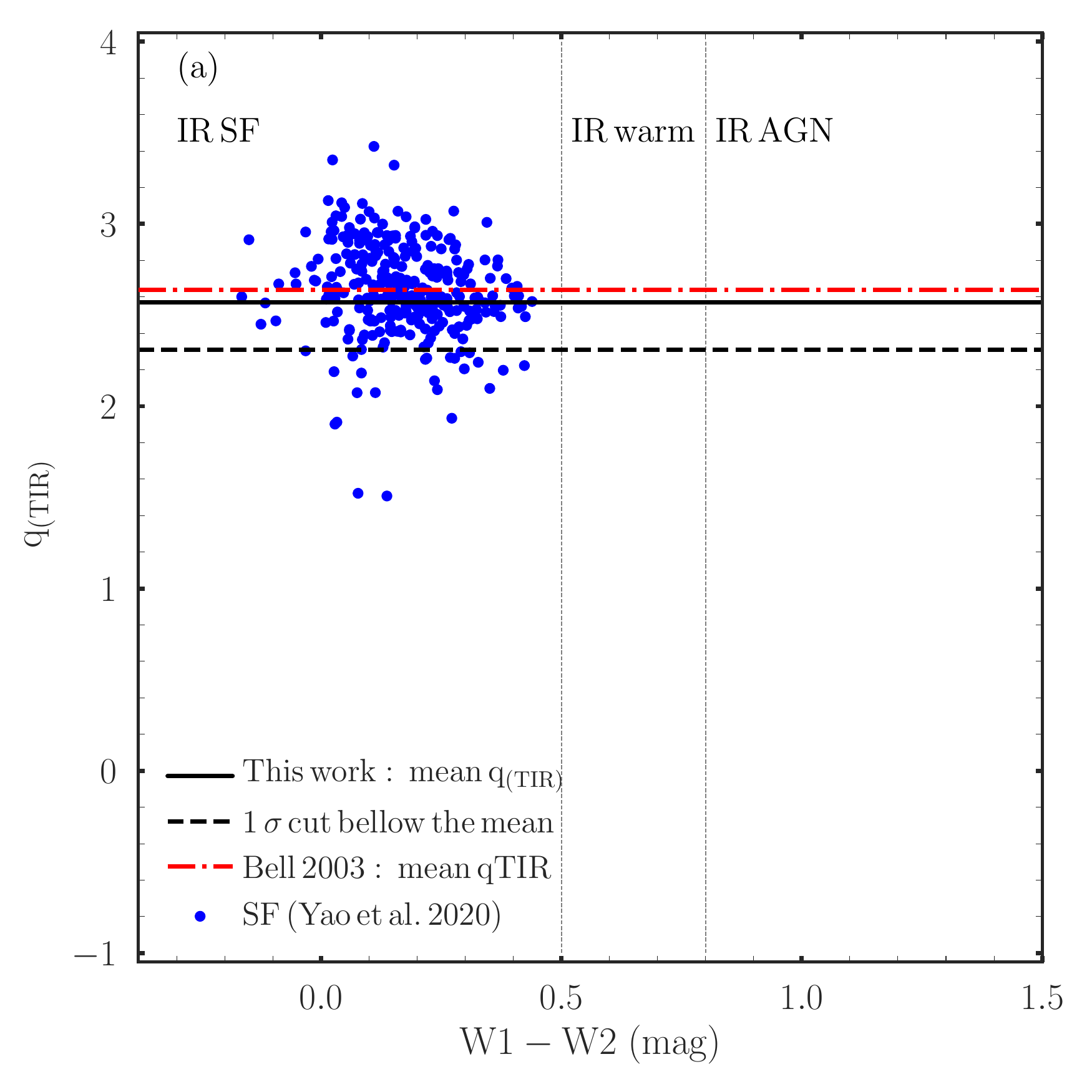}
\includegraphics[width= 8.5cm] {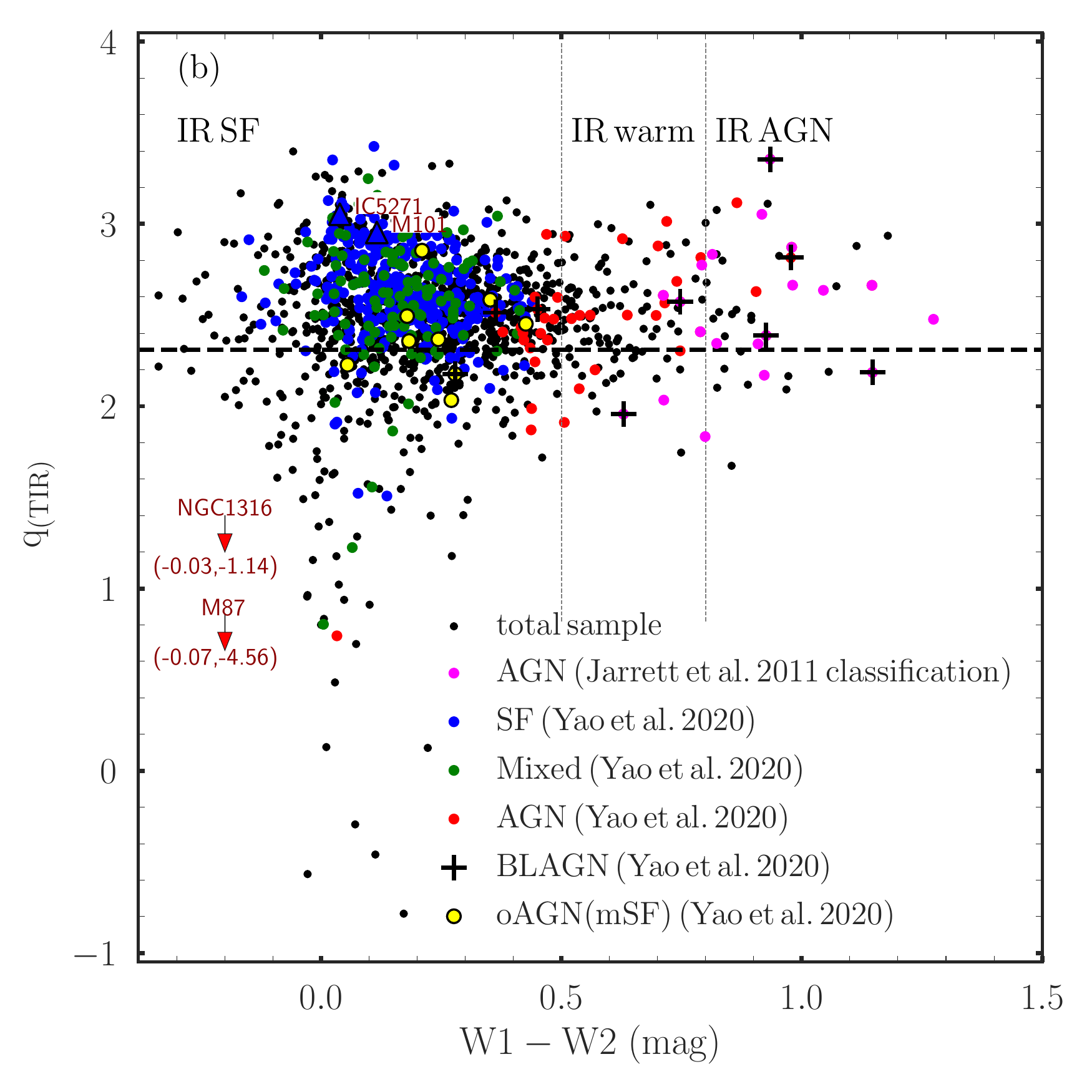}\\ 
\caption{q$_{\text{(TIR)}}$ versus  W1-W2 colors. The W1-W2 color is widely used to separate AGNs from SF galaxies  in infrared  (see Assef et al. \citeyear{Assef2013}) while the \text{q$_{\text{TIR}}$} can discriminate between the radio SF and AGNs.  The blue points represent the SF sample as defined in Section  \ref{New_diagram}.   The horizontal black  and the dashed red lines are our mean  value (mean q$_{\text{(TIR)}}$= 2.57) and that of Bell et al. \citeyear{Bell2003} (q$_{\text{TIR}}=$ 2.64 $\pm$0.02).  The black dashed line is a 1$\sigma$ cut (q$_{\text{(TIR)}}$= 2.31) below the mean q$_{\text{(TIR)}}$. The galaxies above this limit are our most robust SF sample, which will be used in Figures \ref{fig10n} and \ref{fig11}. Our q$_{\text{(TIR)}}$ range from 1.4 to 3.6. All the galaxy samples, similar to Figure \ref{fig7} are represented in (b). The crosses are the broad-line AGNs. The vertical-doted lines are delimitations introduced by Yao et al. (\citeyear{Yao2020}). The SF galaxies are  found at W1-W2 $<$ 0.5 mag. The low power IR AGNs have 0.5$<$W1-W2 $<$ 0.8 mag and the powerfull AGNs (QSOs, Obscured AGNs,  etc.) have W1-W2 $>$ 0.8 mag. 
Two well-known AGNs  host galaxies (M87 and NGC1316) are shown as examples for comparison. They have extremes (upper limit) values given in brackets indicated by the red arrows.  The black points in the background are galaxies of the MeerHOGS--GAMA--WISE sample which do not respect all the criteria (constraints imposed on the S/N of the optical emission lines and \textit{WISE}' colors; see section \ref{MeerHOGS-Cross-matched}) to be classify using either the Yao et al. (\citeyear{Yao2020}) or Jarrett et al. (\citeyear{Jarrett2011}) method.} \label{fig8}
\end{figure*}

In Figure \ref{fig8}a we consider q$_{\text{(TIR)}}$ versus the W1-W2 color, which acts as a mid-IR AGN discriminator (these delineations are from Yao et al. \citeyear{Yao2020}). As already mentioned the  W1 and W2 bands are sensitive to the continuum emission from evolved stars and the hot dust (only W2) making the W1 - W2 color a good diagnostic for AGN emission (see e.g., Jarrett et al. \citeyear{Jarrett2011}; Stern et al. \citeyear{Stern2012}). In panel (a) we show only the SF sample using the classification from Paper I (which is limited to $9 < \log_{10}{\text{stellar}}/\text{M}_{\odot}) < 11$, see next section), which is by definition found in the IR SF zone.

For this SF sample, we calculate a mean value of q$_{\text{(TIR)}}$= 2.57\,$\pm$\,0.23 (black horizontal line). In a sample of 162 SF galaxies,  Bell (\citeyear{Bell2003}) found a median q$_{\text{TIR}}=$ 2.64 $\pm$0.02 (red horizontal line), comparable to our result, albeit with less scatter.A similar result for q$_{\text{TIR}}$ was found by Ocran et al. (\citeyear{Ocran2020}) using data from the Giant Metrewave Radio Telescope (GMRT). They observed an evolution of the  q$_{\text{TIR}}$ with redshift in which the average q$_{\text{TIR}}$ value at low redshift was $\approx$ 2.8. 

The black dashed line in Figure \ref{fig8}a is a 1$\sigma$ cut (q$_{\text{(TIR)}}$= 2.31) below the mean q$_{\text{(TIR)}}$ of our distribution, which we consider to be a more robust SF sample since the radio and infrared emission are consistent, and will be used in Figures 10 and 11.

We investigate simultaneously the optical, infrared, and radio properties of our sample by observing the different AGN-SF categories in the \text{q$_{\text{(TIR)}}$} vs W1-W2 plane (see Figure \ref{fig8}b). All the types of galaxies appear to have comparable \text{q$_{\text{(TIR)}}$} values  -- even the AGNs selected using the selection by Jarrett et al. (\citeyear{Jarrett2011}) and the BLAGN from our sample have similar \text{q$_{\text{(TIR)}}$} compared to the galaxies classified as $\textit{Mixed}$ and SF. Several galaxies in the radio-bright regime are classified as either SFs or AGNs by the Paper I classification scheme (see blue and red points below \text{q$_{\text{(TIR)}}$} $=$ 2.31 in Figure \ref{fig8}b). These represent galaxies whose IR emission is still significant enough to be classified as either SFs or AGNs in $\textit{WISE}$, but with much stronger radio emission than what is expected from the radio-IR correlation (see Eq. \ref{eq6.2}).

Both M\,87 and NGC\,1316 are two well-known AGNs with massive stellar hosts (Jarrett et al. \citeyear{Jarrett2019}) located in nearby galaxy clusters and not from our data set. They have radio fluxes of 138.487\,Jy and 125\,Jy (radio flux from the NVSS data, Condon et al. \citeyear{Condon1998}), respectively. These radio fluxes are extremely high, but their \text{SFR$_{ {12\mu}\text{m}}$} show that they are passive galaxies (SFR[\text{M$_{\odot}$}\text{yr$^{-1}$}] = 0.245 and 0.904 for M\,87 and NGC\,1316, respectively), which have stopped forming new stars. 
On the other hand IC\,5271 is a nearby active SF spiral galaxy with a \text{SFR$_{ {12\mu}\text{m}}$} of 1.64\,\text{M$_{\odot}$}\text{yr$^{-1}$}. It has by far the largest (118$^{\prime \prime}\times$70$^{\prime \prime}$) radio disk in our entire sample (see Figure \ref{fig6.14}).  The second example SF galaxy (M101) is not in our sample, but is a well-known nearby spiral with an active \text{SFR$_{ {12\mu}\text{m}}$} of 3.9\,\text{M$_{\odot}$}\text{yr$^{-1}$} (Jarrett et al. \citeyear{Jarrett2019}).

\begin{figure*}
\centering
\includegraphics[width= 8.5cm] {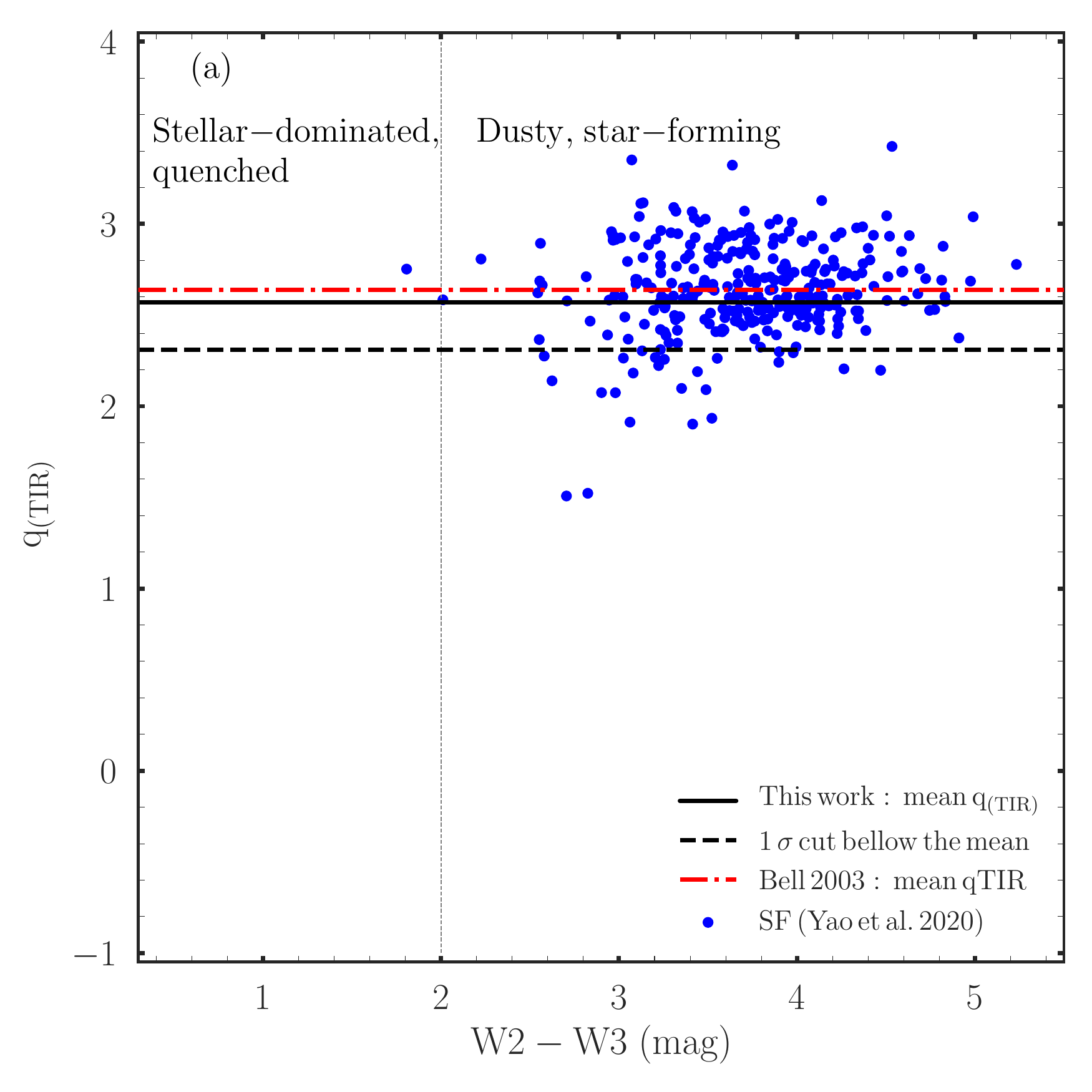}
\includegraphics[width= 8.5cm] {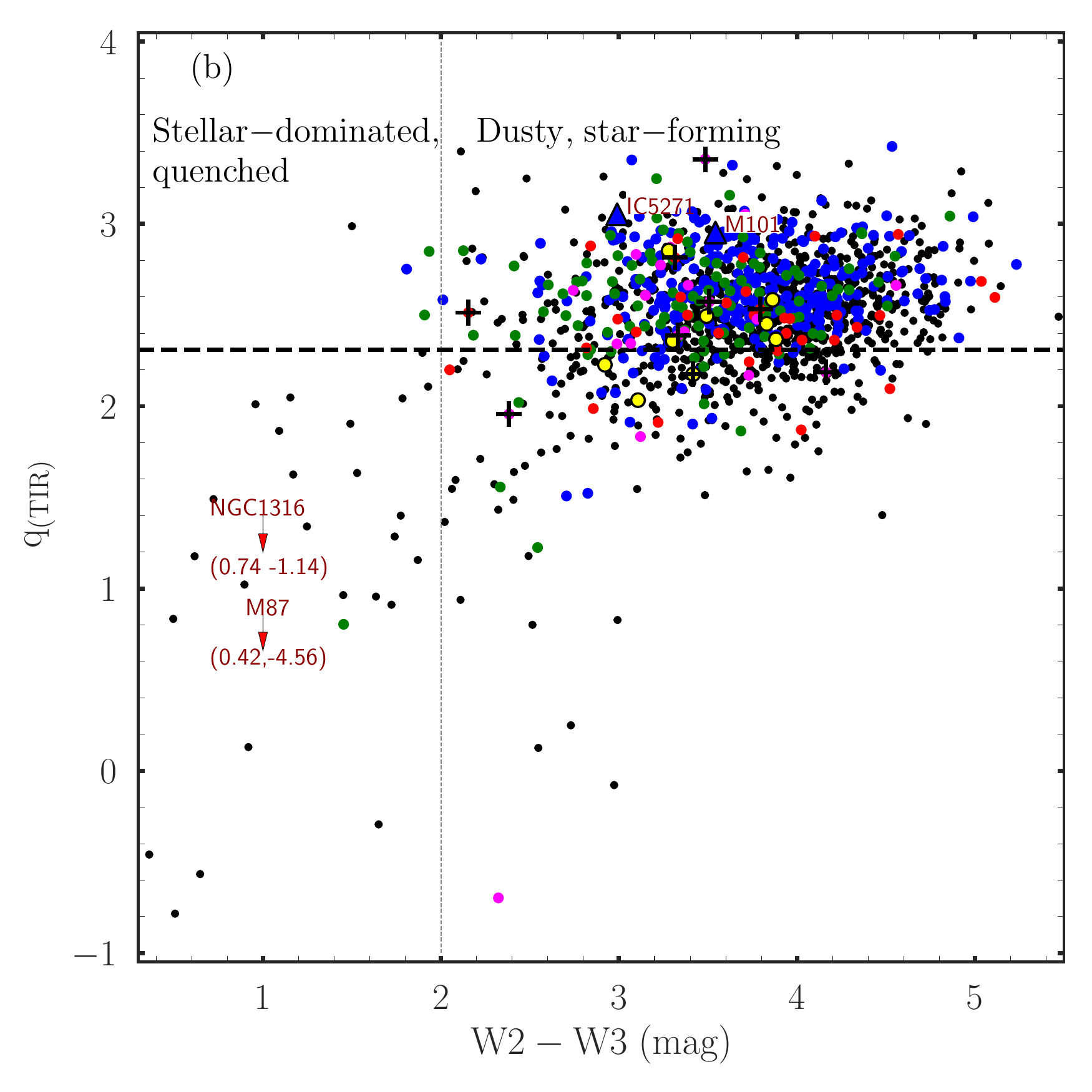}\\ 
\caption{The same groups of galaxies from Figure \ref{fig8} are represented in the q$_{\text{(TIR)}}$ versus  W2-W3 plane. The W2-W3 color mainly divides our sample of  galaxies  into quenched, stellar-dominated for color magnitudes $<$ 2\,mag, and dusty star-forming for higher (redder) magnitudes.
}
\end{figure*} \label{fig9m}

In Paper I we designated galaxies classified as AGN in the optical (BPT) and SF in the IR ($\textit{WISE}$) as ``oAGN (mSF)" and presented them as an interesting class of galaxies to be followed up in radio emission. The aim is to understand why they show strong emission lines, characteristic of AGN in the optical, but low values of W1$-$W2 color.  Nine galaxies from the original IR-optical sample of oAGN (mSF) have radio counterparts (see Table 1 from Paper I). Figure \ref{fig8}b shows these sources (in yellow); they all have q$_{\text{(TIR)}}$ values between two and three, sharing a similar IR/radio luminosity ratio with normal SF galaxies. With their additional properties in radio, we could think of the discordant classification between the IR and the optical of these galaxies as being caused by the dominant emission from the host in the mid-infrared. We could also question the origin of the optical emission lines, chiefly the  [\ion{O}{iii}] lines that can be generated by shocks rather than central AGNs (see Berney et al \citeyear{Berney2015}).\\

The W2-W3 color in Figure \ref{fig9m} can be used to split our sample into  stellar-dominated (generally quenched) and dusty star-forming galaxies at low and high color magnitude, respectively. As In Figure 8, we show only the SF sample in (a) and the full sample in (b). Unlike  the W1$-$W2  color,  the W2$-$W3 color is unable to distinguish between SF galaxies and AGNs as reflected by the  flat distribution for high q$_{\text{(TIR)}}$ values (Figure \ref{fig9m}b), where all categories of galaxies are mixed together. Nevertheless, through the tail of the distribution along the y-axis (also seen in Figure \ref{fig8}b for W1- W2 values close to 0)  we see more radio AGNs moving from dusty and/or actively star-forming galaxies to quenched and apparently inactive (in the infrared) galaxies.  These galaxies with q$_{\text{(TIR)}}$ values  lower than 2.31  and hence anomalously high radio luminosity relative to the IR, will be referred to as ``radio bright galaxies" throughout the rest of this paper.

\subsection{The Variation of \text{SFR$_{ {12\mu}\text{m}}$} and q$_{\text{(TIR)}}$ with Radio Power}

We next explore the radio properties of our SF galaxies. As noted in Paper I, a mass constraint of 9\,$<$\,log$_{10}$\,M$_{\text{stellar}}$\,$<$\,11 has been applied to avoid the metallicity and color degeneracies that affect low mass galaxies (Yao et al. \citeyear{Yao2020}, see Section 3.4). Beyond this mass regime, galaxies are moving off the star-forming main sequence of galaxies (see Cluver et al. \citeyear{Cluver2020}), likely connected to increased AGN activity. 

\begin{figure}[!ht]
\centering
\includegraphics[width=  8.5cm]{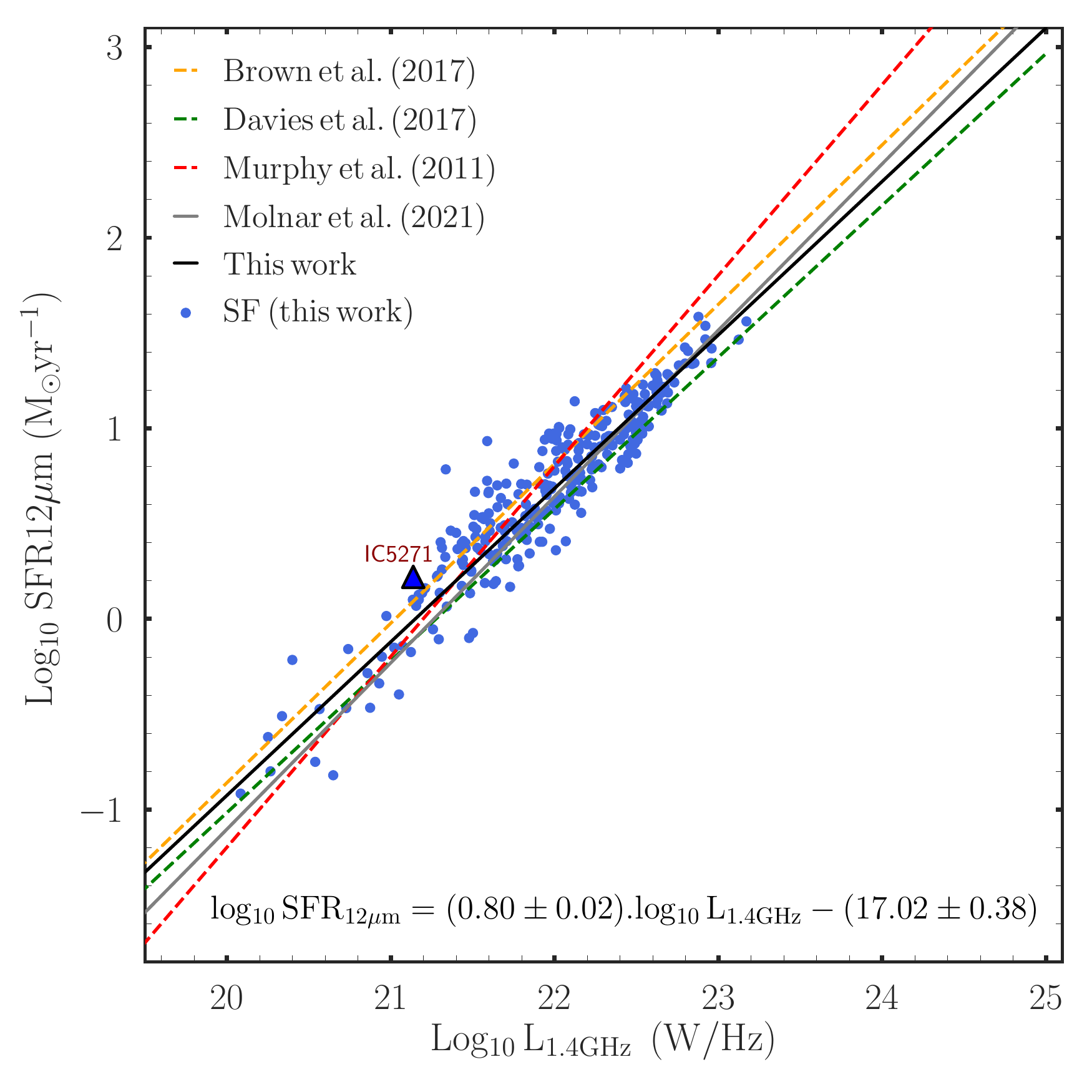}
\caption{The radio luminosity as a function of \text{SFR$_{ {12\mu}\text{m}}$}  for SF galaxies (see SF galaxies with q$_{\text{(TIR)}}$ $>$ 2.31 in Figure \ref{fig8}). The relations of Brown et al. (\citeyear{Brown2017}), Davies et al. (\citeyear{Davies2017}), Murphy et al. (\citeyear{Murphy2011}, radio) and Moln{\'a}r et al. (\citeyear{Molnar2021})  have been added. All the relations are converted to a Kroupa initial mass function (IMF) and the best fit of our data,  given by the equation on the figure is slightly flatter than the others. The radio SFR by Murphy predicts the highest radio values. The blue triangle is the low-$\textit{z}$ (0.00636), high S/N galaxy IC5271 (see Figure \ref{fig6.14}). It is the largest (118$^{\prime \prime}\times$70$^{\prime \prime}$) radio galaxy in our field and is also well resolved in IR. The radio and IR images of IC5271 are represented in  Figures  \ref{fig4} and  \ref{fig6.14}.}\label{fig10n}
\end{figure}

With the addition of radio continuum data, we can now apply a further constraint to construct a more robust SF sample, achieved by applying the additional \text{q$_{\text{(TIR)}}$} $>$ 2.31 cut. This selection removes any likely radio AGNs (see Figure \ref{fig8}) and any potential contamination to our SF sample. In Figure \ref{fig10n} we present this sample with SFR$_{ {12\mu}\text{m}}$ (using the relation from Cluver et al. 2017) shown as a function of radio luminosity.

We include the best fit relations from Brown et al. (\citeyear{Brown2017}), Davies et al. (\citeyear{Davies2017}),  Murphy et al. (\citeyear{Murphy2011}) and Moln{\'a}r et al. (\citeyear{Molnar2021}), rescaled where appropriate to reflect a Kroupa IMF. The data shows the expected increase in the radio luminosity with \text{SFR$_{ {12\mu}\text{m}}$}. The best fit is given by the following relation:

\begin{multline}  \label{eq6.3}
\text{log$_{10}$}\,\text{SFR$_{ {12\mu}\text{m}}$}\,(\text{M$_{\odot}$}\text{yr$^{-1}$})  = (0.80\pm0.02)\times\,\text{log$_{10}$}\,\text{L$_{\text{1.4GHz}}$}\\
\,(\text{W\,Hz}^{-1}) - (17.02\pm0.38)\\
\end{multline}

Our best fit is virtually identical to the fit of Davies et al. (\citeyear{Davies2017}). The difference in Murphy et al. (2011)'s much steeper relation is likely due to the fact their relation is derived using  H$\alpha$ data from diverse SF regions in a single galaxy, as opposed to global flux in a wider range of galaxies used in the other studies, including ours.

\begin{figure}
\centering
\includegraphics[width=  9cm] {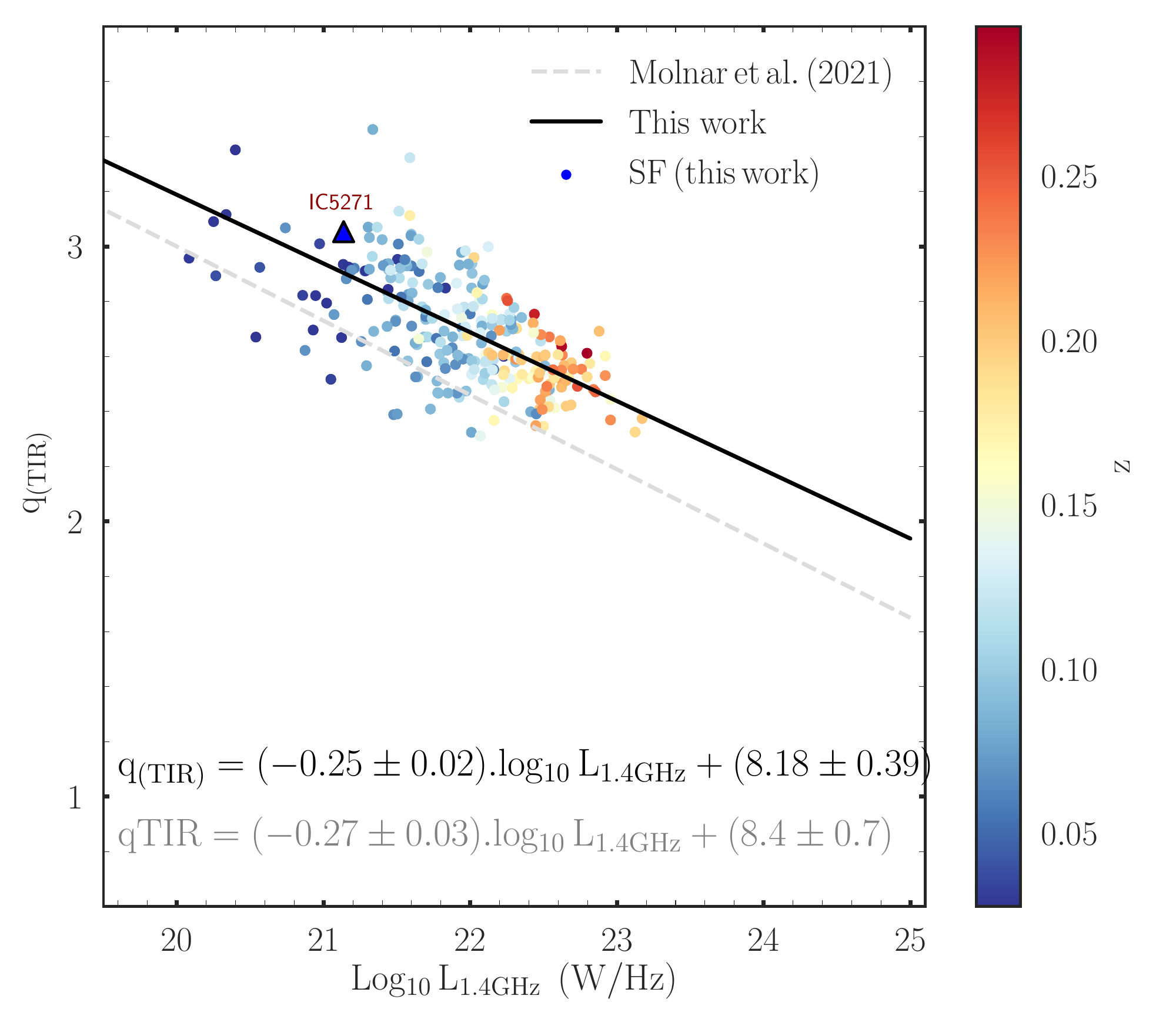}\\ 
\caption{The radio luminosity as a function of  q$_{\text{(TIR)}}$ color-coded by the redshift ($\textit{z}$),  for SF galaxies with  9\,$<$\,log$_{10}$\,M$_{\text{stellar}}$\,$<$\,11 and \text{q$_{\text{(TIR)}}$} $>$ 2.31 
(the mean q$_{\text{(TIR)}}$ of this sample is 2.69 $\pm$0.2). We used the IR radio luminosity  relation as  defined by Helou et al. (\citeyear{Helou1985}). In this relation, the total IR luminosity (L$_{\text{(TIR)}}$) is derived using the equation from Cluver et al. (\citeyear{Cluver2017}) that relates W3\,(12\,$\mu$m) to the  \text{L$_{\text{(TIR)}}$}. The relation derived by Moln{\'a}r et al. (\citeyear{Molnar2021}) is added for comparison.  Our best fit is represented by the black line (see the equation in the  figure).
The overall trend with redshift is likely the result of a distance-luminosity bias in the sample.
%Its slope is very similar  to that of Moln{\'a}r using their depth-matched  SF galaxies, but the trend seen is likely to be a distance-luminosity biais in the sample}. 
The blue triangle is Galaxy IC\,5271, which has the largest  radio disk in our sample. It is an active SF galaxy with  \text{SFR$_{ {12\mu}\text{m}}$} $\sim$ 1.7\,\text{M$_{\odot}$}\text{yr$^{-1}$}.}  \label{fig11}

\end{figure}

\color{black}
In Figure \ref{fig11} we plot \text{q$_{\text{(TIR)}}$} as a function of radio luminosity, color-coded by the redshift, to investigate the relative variation of the two parameters from low-$\textit{z}$ to 0.3, the imposed cut-off.
 We find a decreasing trend in q$_{\text{(TIR)}}$ with increasing radio luminosity for galaxies located at higher redshift.  The trend is unlikely to be an evolutionary effect, instead reflecting a Malmquist bias in the sample, which is incomplete to SF galaxies at higher redshifts. This trend is given by:

\begin{equation}   \label{eq6.4}
\text{q$_{\text{(TIR)}}$}  = (-0.25\pm0.02)\text{log$_{10}$}\,\text{L$_{\text{1.4GHz}}$}\,(\text{W\,Hz}^{-1}) + (8.18\pm0.39)
\end{equation}
for our sample of low $\textit{z}$ ($<$0.3) galaxies.  We note that this is also reported in Moln{\'a}r et al. (\citeyear{Molnar2021}), although their relation is offset to lower q value by 0.21.

It is important to mention that the 10-15\% scatter observed in the q value (see figures 8-11) is a combination of intrinsic (physical) differences and photometric uncertainties, but also scatter that is induced by the rest-frame correction itself, in both the radio and infrared fluxes.  In the radio, the adopted spectral index is a mean value for star-forming galaxies, but will certainly vary from one galaxy to the next.  In the infrared, the k-correction is more complex, taking into account spectral features that redshift into the band-passes.  In Appendix B, we discuss the method used to correct the WISE fluxes to the rest-frame, and consider the expected uncertainties that is contributing to the scatter observed in the q ratio.  For the redshift range of this study, the expected uncertainty in the rest-frame W3 flux and hence luminosity is 5 to 10\% (and lower still for redshifts $<$ 0.1; see Fig. 19).

\subsection{\text{The Relationship between  q$_{\text{(TIR)}}$}, $\textit{WISE}$ Color, and Stellar Mass}\label{Relationship}

\begin{figure}[!h]
\centering
\includegraphics[width=  9cm] {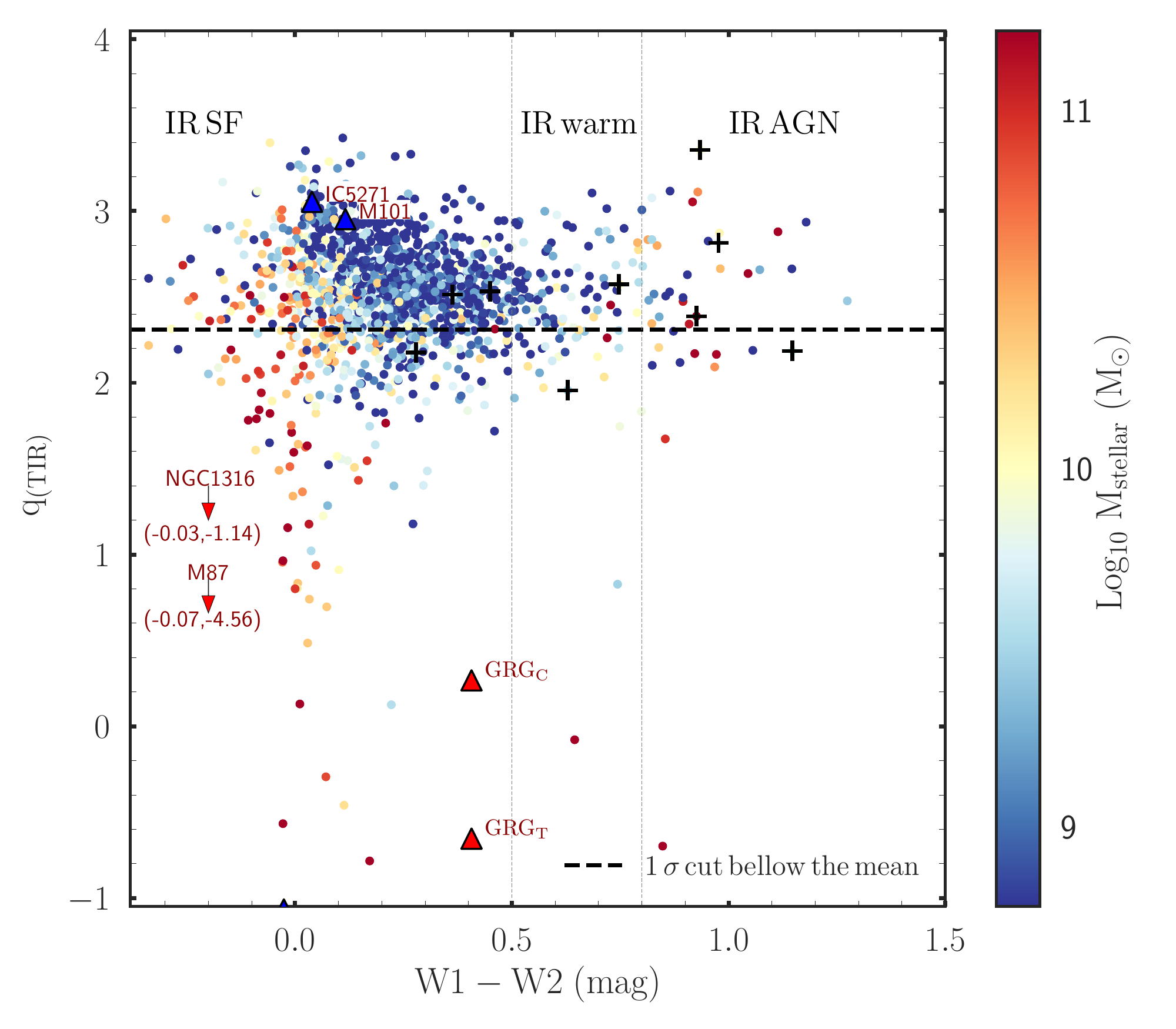} 
\caption{The q$_{\text{(TIR)}}$  vs W1-W2  for the MeerHOGS-$\textit{WISE}$-GAMA sample color-coded as a function of stellar mass (M$_{\text{stellar}}$). The tail, seen in both W1-W2 and W2-W3 colors in the preceding figures (Figures \ref{fig8} and  \ref{fig9m}), appears to be made  of high-mass galaxies. The majority of the highest masses are seen  at very low W1-W2, indicative of stellar colors,  and few cases at high W1-W2 color where  the IR AGNs (including BL specimens) reside. The triangles are galaxies used as examples for comparison. The arrows are used for the galaxies with extreme (upper limit) values. Their names and positions are provided on the graph.}  \label{fig12}
\end{figure}

\color{black}
In Figure \ref{fig12}  the \text{q$_{\text{(TIR)}}$} ratio is shown vs the W1-W2 color and is color-coded by stellar mass. 
While Figures \ref{fig8}b and  \ref{fig9m}b use the radio galaxies only classifiable by the classification method in Paper I (478  galaxies),  Figures \ref{fig12} and \ref{fig13} present all galaxies in the MeerHOGS-$\textit{WISE}$-GAMA sample which can be classified according to $\textit{WISE}$ (1437  from the total sample 1841 galaxies).

The galaxies with high stellar masses commonly have a low W1-W2 color because they are old, quenched, and stellar-dominated; the radio-bright population lies largely in this category. Some high-mass galaxies can be found in the IR AGN region (high W1$-$W2 color).

%Both follow the distribution of the SF galaxies (in blue) well. 
%\color{black}

The GRG (MH5), classified as an AGN in the literature (Seymour et al. \citeyear{Seymour2020}), is in the $\textit{WISE}$ mid-IR ``warm" zone of the color-color diagram (see Figure \ref{fig6.15} for more detail about MH5) and is also powered by its central engine.    We note that the infrared emission is dominated by the AGN and as such the log$_{10}$\,M$_{\text{stellar}}$ is likely slightly overestimated due to hot dust contaminating the W1 emission. This source is a good example of those radio bright AGNs for which the IR emission is strong, but the radio luminosity is very high, hence the low q$_{\text{(TIR)}}$ is such that both the $\textit{WISE}$ color-color diagram and the q$_{\text{(TIR)}}$ classify them as AGNs. Details about the GRG are available in Table \ref{table3} and Figure \ref{fig4}.  %(see the GRG's luminosity in Figure \ref{fig10n})

We therefore consider the $\textit{WISE}$ color-color diagram, color-coded by \text{q$_{\text{(TIR)}}$} and stellar mass, respectively, in panels (a) and (b) of Figure \ref{fig13}. 
\color{black} The stellar-dominated region is where massive ellipticals are believed to have exhausted their reservoirs of cold gas, and are seen as inactive giant galaxies. The \text{q$_{\text{(TIR)}}$} ratio (Figure \ref{fig13}a) reveals a different picture in which although the galaxies appear quiescent in the IR (old star-dominated colors), they are far more active in radio  emission. The radio emission is likely coming exclusively from the central SMBH. M87 is a prototypical example, and was the target of the first-ever image captured of a black hole (Event Horizon Telescope Collaboration \citeyear{Event2019}).

The massive and radio-bright (radio loud or luminous, see Figure \ref{fig13}b) galaxies located at low W2$-$W3 ($<$ 2\,mag), i.e. with very low or imperceptible star formation  activity (SFR\,$\ll$\,1) in $\textit{WISE}$ are a category of AGNs that can only be identified using a combination of radio and IR emission (they generally show absorption lines in optical spectra).

%This comes as an interesting result in a grander picture of trying to identify and study the properties of AGN and SF galaxies using multiwavelength tracers. 
Conversely, the galaxies in the $\textit{WISE}$ warm zone, like that of the QSOs and obscured AGNs, show the same  q$_{\text{(TIR)}}$ range (2\,$<$\,q$_{\text{(TIR)}}$\,$<\,$3.4, see Figure  \ref{fig9m}b) as the normal SF galaxies.  
This shows that  $\textit{WISE}$ is more efficient in detecting the thermal change (hot and warm dust) in a galaxy with characteristics of AGN activity.

\section{Discussion} \label{sec5}

Our observations at $\sim$1.4GHz using the full MeerKAT array found $\sim$18\,000  radio sources in a total area of $\sim$ 10\,deg$^2$  down to a flux density limit of 0.2 mJy. This represents a source density of $\sim$ 1\,800 galaxies/deg$^{2}$. A sample of 1\,841 galaxies forms the combined multi-wavelength sample consisting of radio galaxies having $\textit{WISE}$ and GAMA optical counterparts with W1 $<$ 15.5\,mag (195\,$\mu$Jy).

We carried out the determination of the \text{q$_{\text{(TIR)}}$} ratio for the SF galaxies at redshift $\textit{z}$ $<$ 0.25, which yielded  a range from 2 to 3.4 with a mean value of  2.57\,$\pm$\,0.23, similar to 2.64\,$\pm$\,0.02  found by Bell (\citeyear{Bell2003}). However, there is a clear bias toward lower \text{q$_{\text{(TIR)}}$} values at higher $\textit{z}$ (see Figure \ref{fig11}). 
The range of \text{q$_{\text{(TIR)}}$} from 2.10 to 3.11 in Bressan et al. (\citeyear{Bressan2002}) using 26 compact ULIRGs is also in good agreement with our finding.

The galaxies classified as AGNs (e.g. see classification presented in Figure \ref{fig9m}b) in our sample are not noticeably different from the SF galaxies in terms of the \text{q$_{\text{(TIR)}}$} values. The AGN in these galaxies is definitely affecting the estimation of both luminosities (radio and IR) such that the \text{q$_{\text{(TIR)}}$} is still deceptively consistent  with that of SF galaxies. AGNs will typically be found at \text{q$_{\text{(FIR)}}$} $<$ 2 (Baan $\&$ Klockner  \citeyear{Baan2006}) when the galaxy is no longer forming stars.

Both types of AGNs (red and magenta; Yao et al. \citeyear{Yao2020} and Jarrett et al. \citeyear{Jarrett2011}) share similar  \text{q$_{\text{(TIR)}}$} values to the SF galaxies (blue) as presented in Figure \ref{fig8}b.  However, their W1-W2 colors indicates a hot dust component, likely associated with AGN accretion, which confirms AGN activity that clearly separates them from pure SF galaxies in this diagram. Hence, in this case, the radio luminosity and the W1$-$W2 color reveals the presence of AGN activity.

%\newpage
\thispagestyle{plain} 
\begin{figure*}%[!h]
\vspace*{-0.8in}
\centering
\includegraphics[width=  12.9cm] {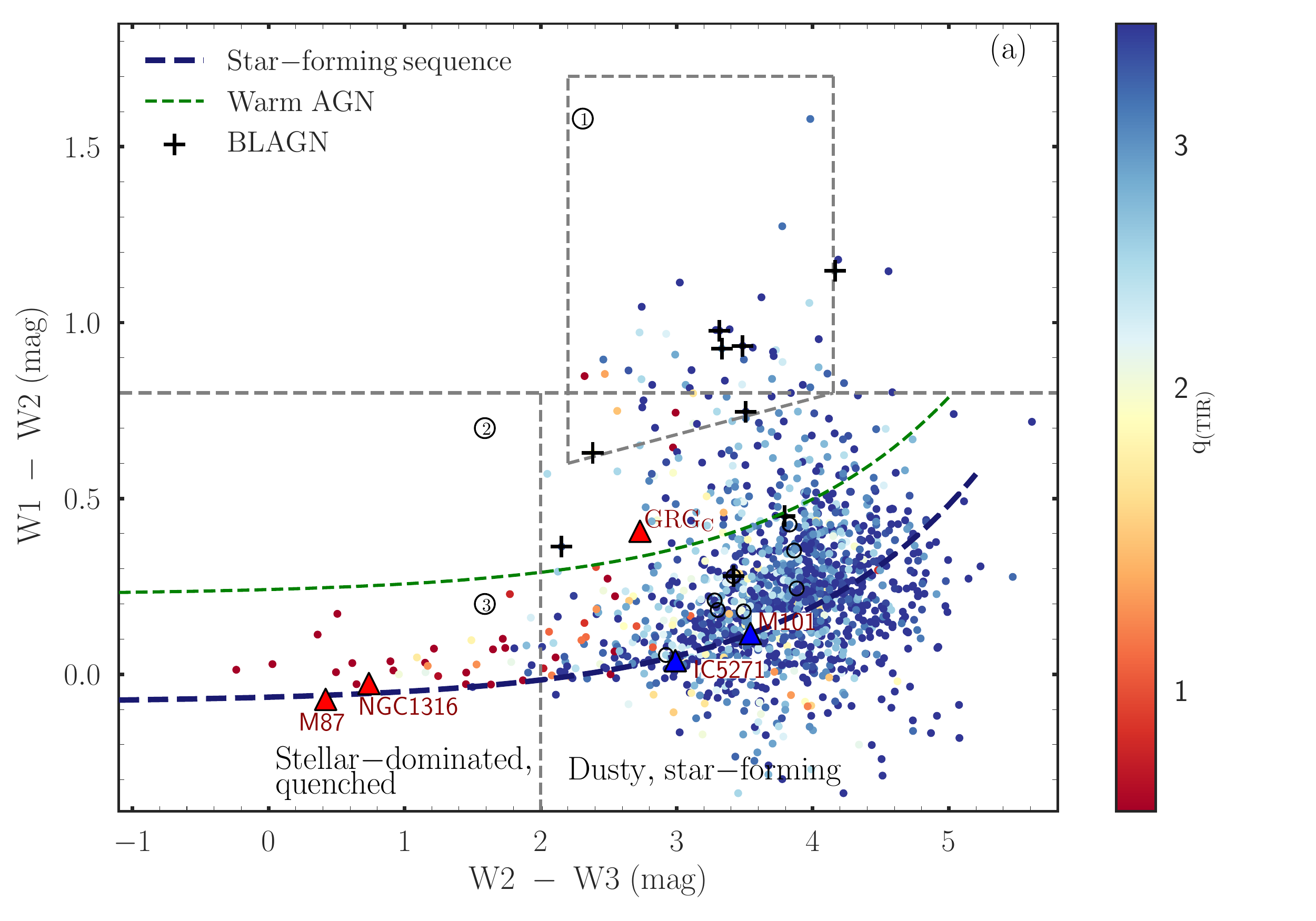}\\ 
\includegraphics[width=  12.9cm] {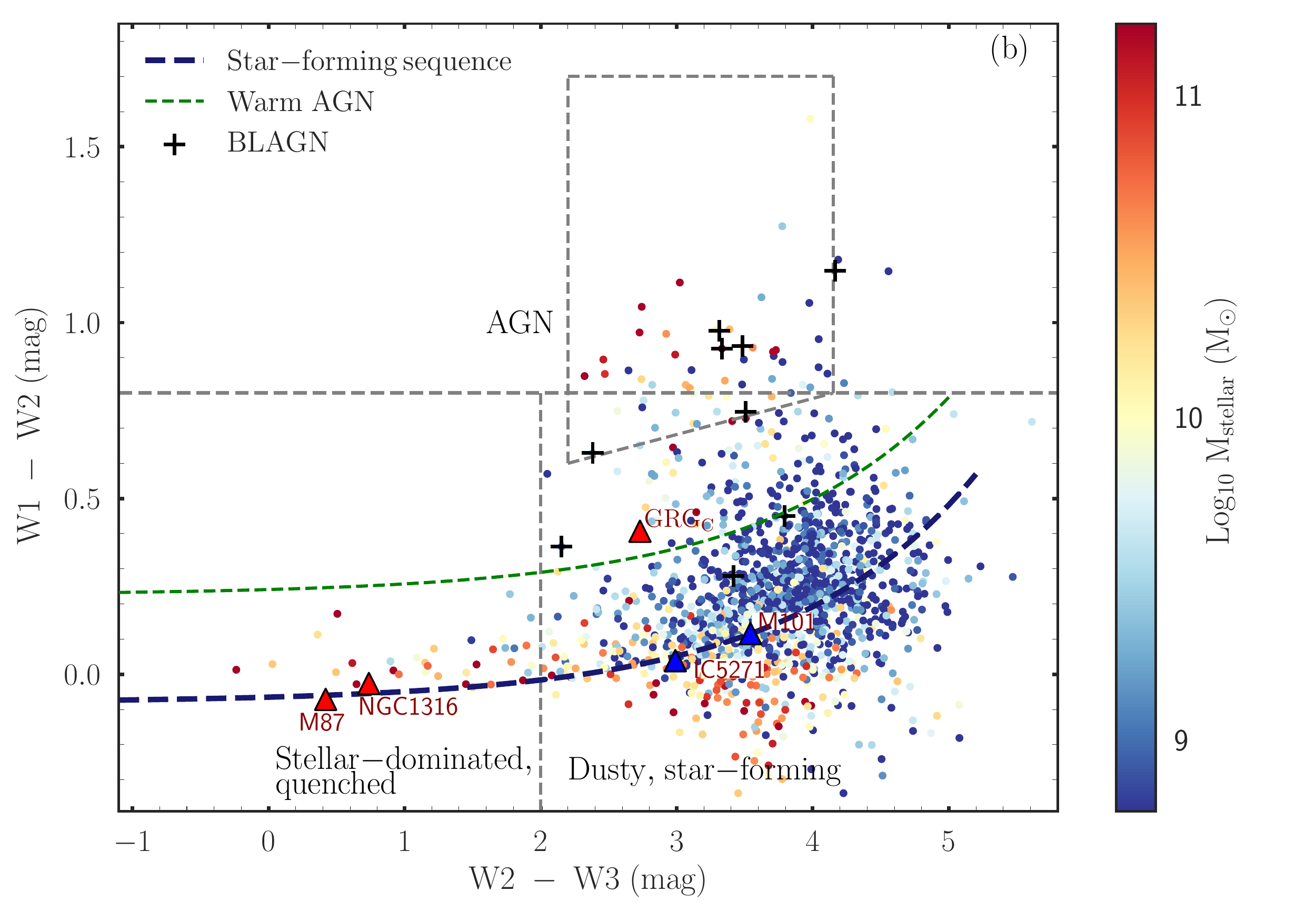}\\ 
\caption{The $\textit{WISE}$ color-color diagram coded by the  q$_{\text{(TIR)}}$ and the stellar mass in (a) and (b), respectively. The Zone 1, 2 and 3 in (a) are the respective locations of the $\textit{WISE}$ powerful AGN (Jarrett et al. \citeyear{Jarrett2011} classification), the low power AGN and finally the non-AGN zone mostly populated by star-forming  and quenched galaxies (see Figure 9 in Yao et al. \citeyear{Yao2020}, here the spheroids are included in region 3).  We can see   the galaxies with the lowest q$_{\text{(TIR)}}$ in the quenched and stellar dominated zone. It is quite interesting to see how the galaxies in this area of the $\textit{WISE}$ color-color diagram appear to be very prominent in radio. Since star formation is mostly associated with dust, not having  a dust signature (very low SFR) in $\textit{WISE}$ means that the radio flux is probably coming exclusively from the central AGN (local examples are M87 and NGC 1316). This comes as a complementary finding in our quest to identify all forms of AGN through multiwavelength study. GRG$_{\text{C}}$  represents the core (41.4 mJy) of the giant radio galaxy (PKS 2250--351) from our sample (MH5). The blue and green dashed sequences' relations were first introduced in Figure 9 from Yao et al. (\citeyear{Yao2020}). The galaxies between the green dashed line and the AGN box are classified as $\textit{WISE}$ warm galaxies and are believed to be  low-power AGNs (Yao et al. \citeyear{Yao2020}). However, their q$_{\text{(TIR)}}$ looks similar to that of normal SF galaxies. The black circles are galaxies classified as AGN in the optical BPT, but SF using the infrared colors (see legend of Figure \ref{fig8}b).}  \label{fig13}
\end{figure*}
%\newpage

In our recent study, which combined carefully measured $\textit{WISE}$ photometry and optical data (Yao et al. \citeyear{Yao2020}), we found several galaxies with optical lines characteristic of AGNs, but with  low W1-W2 mag values, indicative of normal star-forming processes. These have been classified as ``optical AGNs ($\textit{WISE}$ SF)" and are particularly interesting to examine in the radio. Nine of these galaxies (GAMA ID: 5154519, 5155115, 5222053, 5234844, 5237886, 5240322, 5240366, 5258350, and 5427366)  have radio counterparts in MeerHOGS. They do not show clear excess in radio and have \text{q$_{\text{(TIR)}}$} values ranging between  2 and 2.8 with a mean of 2.38.
The host likely dominates the IR and radio emission (see 5155115 in Figure 25 from Paper I) in which case luminosities in both wavelength regimes are similarly affected. It could also signify that the optical lines in theses galaxies, although mimicking AGN activity, are not (partially or fully) induced by a central AGN. They exhibit high log L[\ion{O}{iii}]  ($>$\,41.2\,erg/s) and 
10.1$<$\,Log(M$_\text{stellar}$/M$_{\odot}$)\,$<$ 11.0 with a mean of 10.61.  In the current study, the \text{q$_{\text{(TIR)}}$}, like the infrared, does not indicate AGN activity in these galaxies.  It seems to confirm the hypothesis that their accretion disk is not powerful enough to generate the winds that produce broad-line regions  or alternatively that the [\ion{O}{iii}] lines are coming from non-nuclear shocks rather than from the central AGN (see Yao et al. \citeyear{Yao2020} for further discussion).%  

On the other hand,  radio excess was observed in some $\textit{WISE}$ galaxies, mainly among the stellar-dominated, quenched population. The IR--X-ray study by Huang et al. (\citeyear{Huang2017}) showed that building SEDs with data collected at several different frequency bands (combining $\textit{WISE}$, $\textit{Spitzer}$ and $\textit{AKARI}$ that provide continuous IR coverage from 2 to 24 $\mu$m) helps reveal an important fraction of hidden AGNs in SF galaxy samples. 
Indeed, the four $\textit{WISE}$ bands are robust enough to identify AGNs, but are limited in cases where the accretion energy of the AGN (low power AGNs) is submerged by the star formation activity and the underlying stellar population of the host. Including an AGN model in the  SED fits ( e.g., ProSpect; Robotham et al. \citeyear{Robotham2021}) appears to be a viable method, which can then reliably identify these low power AGNs, with the advantage of recovering  AGN luminosities (see Thorne et al. \citeyear{Thorne2022}).   
Another issue in this regard is whether there is enough dusty material around the AGN to be clearly detected and delineated in the IR. These cases are frequent among  AGNs with a ``hot" accretion mode (see review by Fabian \citeyear{Fabian2012}) known as low excitation radio galaxies (LERGs), which emit almost exclusively in the form of powerful radio jets. They are seen in our IR-radio analysis as galaxies having \text{q$_{\text{(TIR)}}$} $<$ 2.31 combined with low SFR$_{ {12\mu}\text{m}}$.

Several studies in the literature (Magnelli et al. \citeyear{Magnelli2015}; Delhaize et al. \citeyear{Delhaize2017}; Calistro Rivera et al. \citeyear{Calistro2017}; Ocran et al. \citeyear{Ocran2020})  have shown a variation of the  q$_{\text{(TIR)}}$ with redshift.  Although our study does not focus on the evolution of the FIR/radio luminosity with redshift, we do see a decrease of the \text{q$_{\text{(TIR)}}$} with the radio luminosity (see Figure \ref{fig11}) similar to the result found in Moln{\'a}r et al. (\citeyear{Molnar2021}). We place our q$_{\text{(TIR)}}$ values in context in Figure \ref{fig11} -- this confirms that the decrease we observe is related to redshift. Indeed, as expected, the radio  luminosities increase with redshift as would be the case for a luminosity selection effect. 
At relatively low redshifts, $\textit{z}$ $<$ 0.15, we find \text{q$_{\text{(TIR)}}$} with a value above 2.5, in agreement with that of Bell (\citeyear{Bell2003}) and Moln{\'a}r et al. (\citeyear{Molnar2021}).

\begin{table*}%[!hb]
\small
\centering
\caption{Classification matrix, table showing the combination of  3 methods for more effective separation of galaxies into AGN and SF classes. The AGNs identified using the high W1-W2 color cut  ($>$ 0.8\,mag) by Jarrett et al. (\citeyear{Jarrett2011}; JC) are also included in the  classification of Yao et al. (\citeyear{Yao2020}; YC), except those for which the optical lines are in absorption.  On the other hand, the YC can identify the low power AGN not taken into account by JC. A significant proportion of the AGNs using the \text{q$_{\text{(TIR)}}$} is not seen by JC and YC. However, the \text{q$_{\text{(TIR)}}$} is not very efficient in the identification of low-power AGNs. It is unlikely for a galaxy to be classified as AGN in JC and not in YC.} \label{summary_table} 
\begin{tabular}{lccccc}
\tablewidth{0pt}
\hline
&Jarrett et al. (\citeyear{Jarrett2011}) class.&Yao et al. (\citeyear{Yao2020}) class.&qTIR class.&\textbf{final  class.}&\textbf{Examples}\\ \hline
&AGN?&AGN?&AGN?&\textbf{AGN?}&\textbf{See galaxies in Fig. \ref{fig13}a} \\
Case 1&yes&yes&yes&\textbf{yes}&Low \text{q$_{\text{(TIR)}}$} ($<$2.3) in region 1 \\
Case 2&yes&yes&no&\textbf{yes}&High \text{q$_{\text{(TIR)}}$} ($>$2.3) in region 1\\
Case 3&no&yes&yes&\textbf{yes}&Low \text{q$_{\text{(TIR)}}$} in region 2\\
Case 4&no&yes&no&\textbf{yes}&High \text{q$_{\text{(TIR)}}$} in region 2\\
Case 5&no&no&yes&\textbf{yes}&Low \text{q$_{\text{(TIR)}}$} in region 3\\ \hline
&Jarrett et al. (\citeyear{Jarrett2011}) class.&Yao et al. (\citeyear{Yao2020}) class.&qTIR class.&\textbf{final  class.}&\textbf{Examples}\\ \hline
&SF or Mixed?&SF or Mixed?&SF or Mixed?&\textbf{SF or Mixed?}&\textbf{See galaxies in Fig. \ref{fig13}a } \\ 
Case 6&yes&yes&yes&\textbf{yes}&High \text{q$_{\text{(TIR)}}$} in region 3\\
Case 7&yes&no&no&\textbf{no}&Low \text{q$_{\text{(TIR)}}$} in region 2\\
Case 8&yes&yes&no&\textbf{no}&Low \text{q$_{\text{(TIR)}}$} in region 3\\
Case 9&yes&no&yes&\textbf{no}&High \text{q$_{\text{(TIR)}}$} in region 2\\
Case 10&no&no&yes&\textbf{no}&High \text{q$_{\text{(TIR)}}$} in region 1\\\hline
\end{tabular}
\end{table*}

In the $WISE$ color-color diagram (Figure \ref{fig13}) the green curve represents the limit between the pure SF systems (with a few exceptions) and galaxies which we believe have some degree of AGN activity. Above this line are the infrared ``warm" galaxies (from Paper I), which could be a mixture of AGN and SF activity. At larger W1$-$W2 color (W1-W2 $>$ 0.8\,mag) lie galaxies with stronger AGN activity in the infrared, which are mainly quasars, broad-line, and obscured AGNs.  But, most of the galaxies in these AGN-dominated regions of the color-color diagram (see Figure \ref{fig13}a) have high values of  \text{q$_{\text{(TIR)}}$} similar to that of SF galaxies. The \text{q$_{\text{(TIR)}}$} is failing to identify most of the AGNs in this region. We think the limitations of the \text{q$_{\text{(TIR)}}$}  is related to the fact that the hot accretion from the AGN is creating mid-infrared emission that  mimics star formation (see Cluver et al. \citeyear{Cluver2017}; Jarrett et al. \citeyear{Jarrett2017}).

 Finally, we present Table \ref{summary_table} as a summary of the classification methods used in Paper I (see Section \ref{New_diagram}), extended in this paper to include \text{q$_{\text{(TIR)}}$}. This serves to show the power (and limitations) when combining optical, mid-infrared and radio diagnostics to efficiently separate AGN- and SF-dominated galaxies. This is therefore a summary of the results of both Paper I and the current paper.

 We recall that the Yao et al. (\citeyear{Yao2020}) classification takes into account both the optical and infrared properties of the galaxies. In general, a galaxy is likely to have AGN activity when at least one of the methods can detect it. It follows that a galaxy has to be classified as a SF system in all three methods to be accepted as such.

 For example,  case 5 or 8 in Table 4 is a scenario where both the Jarrett et al. (\citeyear{Jarrett2011}) and Yao et al. (\citeyear{Yao2020}) classifications, respectively referred to as JC and YC, do not indicate any AGN activity, but the radio power is able to classify a source as an AGN based on the \text{q$_{\text{(TIR)}}$}. Such galaxies radiate the bulk of their flux at radio wavelengths and generally have weak emission in the infrared and/or the optical lines are in absorption. They are, therefore, classified as AGNs, or radio galaxies.  Case 5 (or 8) represents a galaxy in region 3 of Figure \ref{fig13}a, with a low \text{q$_{\text{(TIR)}}$} value ( \text{q$_{\text{(TIR)}}$} $<$ 2.31, characteristic of radio AGNs). Case 9 or 4 is  a SF in JC and AGN in YC (it is therefore a weak AGN located in region 2 of Figure \ref{fig13}a) with a high \text{q$_{\text{(TIR)}}$} ( \text{q$_{\text{(TIR)}}$} $>$ 2.31) characteristic of radio SF galaxies. 
Case  10 is similar to 2, where only the \text{q$_{\text{(TIR)}}$} does not detect the AGN activity. We think such AGNs have very low emissions in radio. On the other hand,  cases 3 or 7 represents a low power AGN in infrared (according to the Yao et al. 2021 Classification), and not  detected using the classical AGN classification method (Jarrett et al. 2011), but its presence is confirmed by the low  \text{q$_{\text{(TIR)}}$ ratio}. Cases 1 and 6 show no  ambiguity since all methods agree on the classification. \\

In this study, we consider the combination of optical spectroscopy, infrared and radio imaging to more reliably identify and separate star-forming and AGN-dominated galaxies, and combinations of the two phases.  In this emerging
era  of large radio surveys such as MeerKAT and ASKAP and soon the SKA itself, it is vital that we have empirical tools for classifying galaxies by their dominant emission mechanisms which belie their evolutionary stage and path forward.   There is no perfect way of doing this classification, there will always be uncertainties and overlap between different populations.  As with the extensive previous work in this field since the days of IRAS when the infrared window provided a new view into the galaxy evolutionary process, it is our aim to mitigate systematic mis-classifications (reliability) and improve upon the sample completeness and source characterization using the latest data available, such as the \textit{WISE} full-sky image survey. Yet it remains a difficult challenge, 
AGNs 
%but We should recall that AGNs 
are very complex structures, whose observed properties depend on factors such as the viewing angle (see Antonucci  \citeyear{Antonucci1993}; Beckmann $\&$ Shrader \citeyear{Beckmann2012}),  Black Hole (BH) mass, the optical depth, the duty cycle, etc.
Generally, the most massive galaxies  have at some point AGN activity due to the commonly
seen supermassive BH at their centre and the best way to classify them as  SF  or AGN activity dominated is to compare the relative energetic importance of the two components (e.g., Kirkpatrick et al. \citeyear{Kirkpatrick2015}).  Our optical-infrared-radio schema delineated in Table 4 attempts to bring some clarity to this critical step.

\section{Conclusions} \label{sec6}

We processed the radio continuum data obtained with MeerKAT (64 dishes) at 1.4 GHz for a total area of $\sim$ 10 deg$^2$ in the southern region of GAMA (G23), known as the MeerHOGS survey. Although the observation time was only 16.5 hours, we obtained excellent image quality down to 20 $\mu$Jy because of the exceptional sensitivity of MeerKAT. We found several notable multicomponent objects, as presented in Figure \ref{fig4} and \ref{fig6.2b}. The radio sources were extracted and position cross-matched to $\textit{WISE}$ galaxies with GAMA redshift counterparts. This IR-optical-radio catalogue was used as a follow-up to the work in Paper I, notably to study the IR/radio luminosity ratio. Our main findings are presented as follows: 

\begin{itemize}

\item[$\bullet$] The q$_{\text{(TIR)}}$ decreases with L$_{\text{1.4GHz}}$ for SF galaxies and its values ranges from 2 to 3.4 with a median value in the local universe ($\textit{z}$\,$<$0.25) of  2.57\,$\pm$\,0.23. We relate this decrease to the Malmquist bias rather than a physical phenomenon.

\item[$\bullet$] We found a tight correlation between the SFR$_{ {12\mu}\text{m}}$ and the L$_{\text{1.4GHz}}$ for  SF galaxies given by:\\ 
log$_{10}$\,SFR$_{ {12\mu}\text{m}}$ = (0.8\,$\pm$\,0.02)\,log$_{10}$\,L$_{\text{1.4GHz}}$ - (17.02\,$\pm$\,0.38)\\
and consistent with previous studies.

\item[$\bullet$]The galaxies identified as stellar-dominated (quenched) in the $\textit{WISE}$ color-color diagram generally have low SFR and q$_{\text{(TIR)}}$ values $<$2.31.  These galaxies, considered to be inactive in the IR are powerful AGNs that probably emit most of their power at radio wavelengths. In this work, we identify them as ``radio-bright".

\item[$\bullet$] The q$_{\text{(TIR)}}$ robustly identifies AGNs where the radio emission dominates because the host galaxy star formation is generally quenched. It becomes less reliable for cases where the galaxy, although showing clear signatures of AGN activity, still has active star formation. These AGNs in most cases share a similar q$_{\text{(TIR)}}$ range compared to that of robustly identified star-forming galaxies.

\item[$\bullet$] The BPT and $\textit{WISE}$ color-color diagrams  are unable to accurately classify these radio bright galaxies due to their lack of distinctive AGN emission at optical and infrared wavelengths. On the other hand the q$_{\text{(TIR)}}$ is limited to radio-bright galaxies. We therefore suggest the combination of the three methods is best for classifying AGNs and SF galaxies. The combination of the q$_{\text{(TIR)}}$ and the WISE color-color (or the optical BPT) diagram is therefore quite powerful. See Table \ref{summary_table} for a summary of the classifications addressed in this work.

\end{itemize}

\section{ACKNOWLEDGEMENTS}

HFMY would like to acknowledge the support given from the NRF (South Africa) through the Centre for Radio Cosmology (CRC). MC is a recipient of an Australian Research Council Future Fellowship (project number FT170100273) funded by the Australian Government. T.H.J. acknowledge support from the National Research Foundation (South Africa).  This publication makes use of data products from the Wide-field Infrared Survey Explorer, which is a joint project of the University of California, Los Angeles, and the Jet Propulsion Laboratory/California Institute of Technology, funded by the National Aeronautics and Space Administration.\\
The MeerKAT telescope is operated by the South African Radio Astronomy Observatory, which is a facility of the National Research Foundation, an agency of the Department of Science and Innovation.\\
GAMA is a joint European-Australasian project based around a spectroscopic campaign using the Anglo-Australian Telescope. The GAMA input catalog is based on data taken from the Sloan Digital Sky Survey and the UKIRT Infrared Deep Sky Survey. Complementary imaging of the GAMA regions is being obtained by a number of independent survey programs including GALEX MIS, VST KIDS, VISTA VIKING, WISE, Herschel- ATLAS, GMRT and ASKAP providing UV to radio coverage. GAMA is funded by the STFC (UK), the ARC (Australia), the AAO, and the participating institutions.  Based on observations made with ESO Telescopes at the La Silla Paranal Observatory under programme ID 177.A-3016. The GAMA website is http://www.gama- survey.org/.

\software{CARACal  (J\'{o}zsa et al.  \citeyear{Jozsa2020}), AOFlagger  (Offringa  \citeyear{Offringa2010}), ProFound  (Robotham et al. \citeyear{Robotham2018}), Stimela  (Makhathini  \citeyear{Makhathini2018}), WCSClean  (Offringa et al.  \citeyear{Offringa2014}),
 CUBICAL  (Kenyon et al.  \citeyear{Kenyon2018}), SoFiA  (Serra et al.  \citeyear{Serra2015}), VSAD  (Condon et al.  \citeyear{Condon1998}), AEGEAN   (Hancock et al.  \citeyear{Hancock2012},  \citeyear{Hancock2018}), pyBDSF  (Mohan  \citeyear{Mohan2015}),  mpfitfun (Markwardt \citeyear{Markwardt2009}).}

\newpage

\appendix

\section{Resolved galaxies}

\begin{figure*}[!hb]
%\hspace*{-0.8in}
\includegraphics[width= 15.5cm] {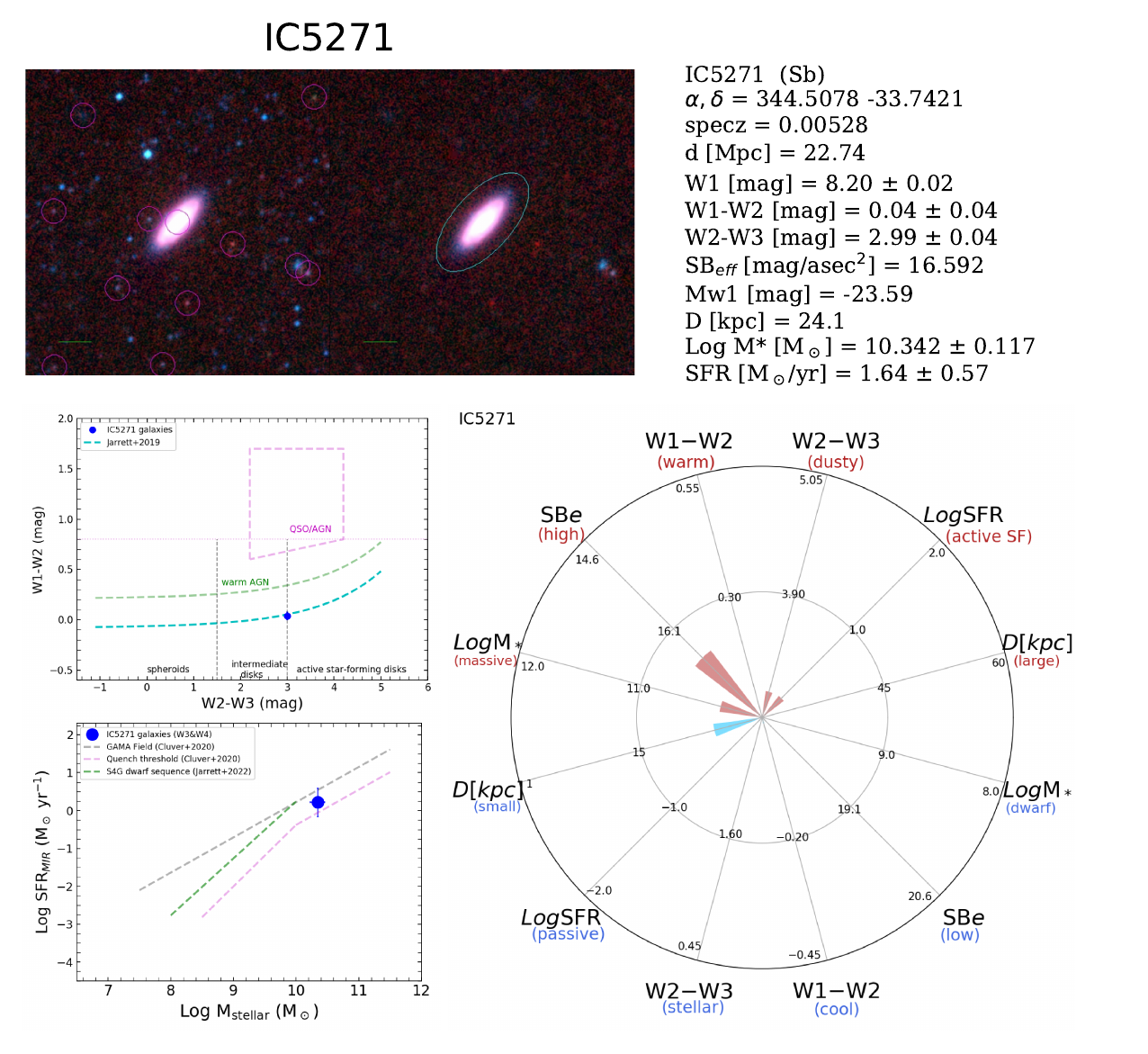} 
\caption{ Physical properties of the galaxies selected as well-known cases. Here we show the infrared properties of IC\,5271, which includes the colors and physical properties.  
The lower right panel shows a pinwheel diagram, which  displays the physical properties  of the galaxies (see Jarrett et al. \citeyear{Jarrett2019} for more details about the diagram). IC5271 (MH4) is classified as SF in $\textit{WISE}$ with a radio flux of 0.0227 Jy.} \label{fig6.14}
\end{figure*}

\begin{figure*}
\centering
%\hspace*{-0.8in}
\includegraphics[width= 16.5cm] {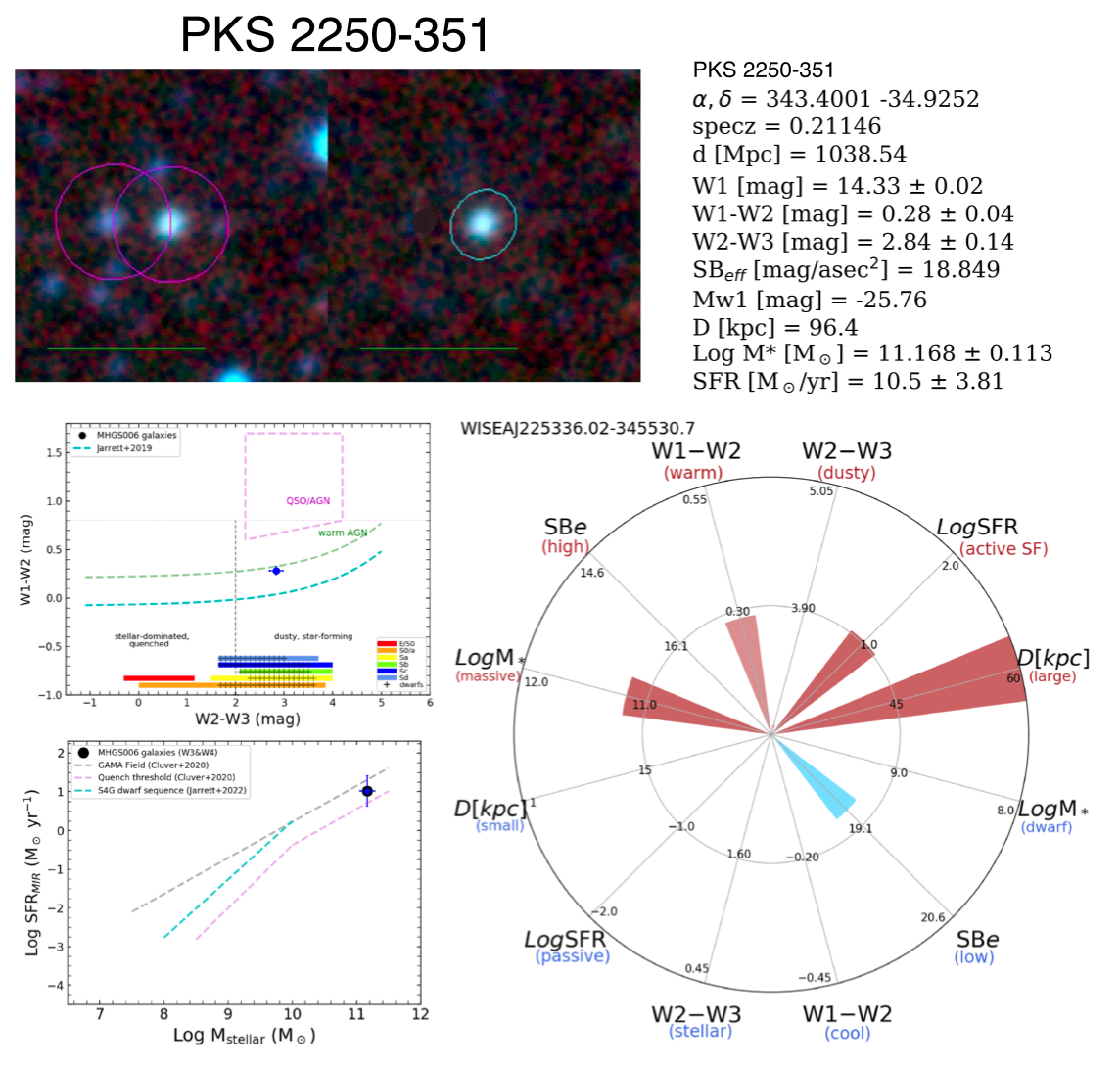} 
\caption{The radio image presented in Figure \ref{fig4}, MH5 is a giant radio galaxy (PKS 2250--351 in Abell 3936, Seymour et al. \citeyear{Seymour2020})  classified as SF in $\textit{WISE}$ with a MeerKAT radio flux of 0.318 Jy.}  \label{fig6.15}
\end{figure*}

Figures \ref{fig6.14} and \ref{fig6.15} represent galaxies  MH4 and MH5, respectively  from Figure \ref{fig4}. They are also used as case examples throughout the document.  The $\textit{WISE}$ 3-color image and the color-color along with the main sequence diagrams are presented. The pinwheel (Jarrett et al. \citeyear{Jarrett2019}) gives a good summary of all the properties such as colors,  stellar mass or SFR etc. 

\begin{figure*}[!ht]
\centering
%\hspace*{-0.4in}
\includegraphics[width= 3.9cm] {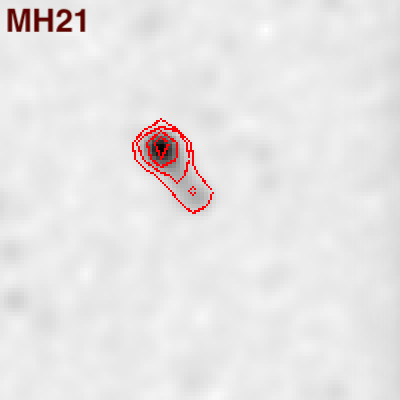}
\includegraphics[width= 3.9cm] {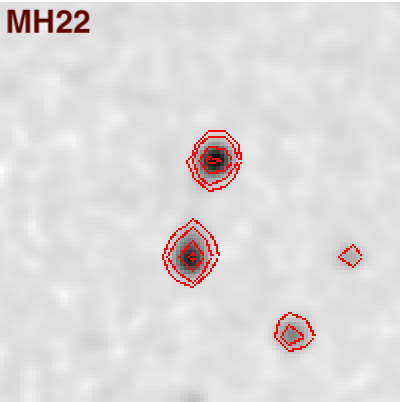} 
\includegraphics[width= 3.9cm] {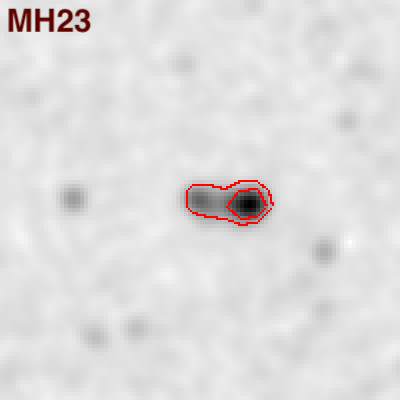} 
\includegraphics[width= 3.9cm] {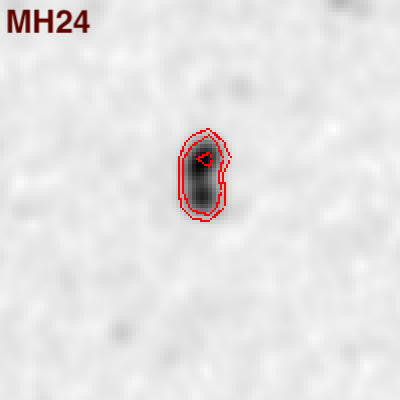}\\
%\hspace*{-0.4in}
\includegraphics[width= 3.9cm] {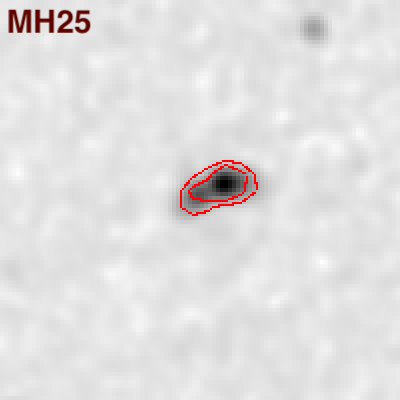} 
\includegraphics[width= 3.9cm] {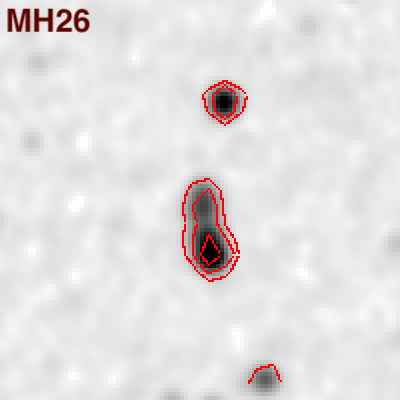} 
\includegraphics[width= 3.9cm] {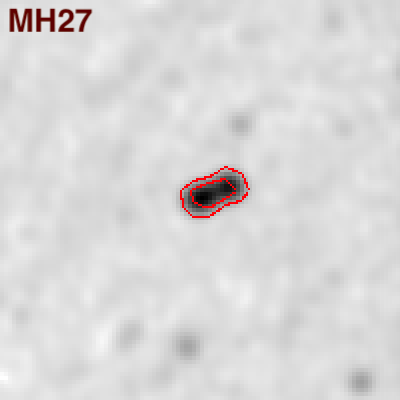} 
\includegraphics[width= 3.9cm] {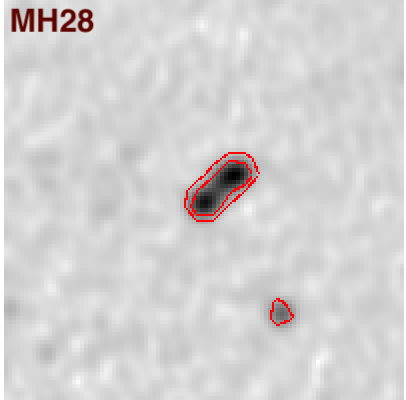}\\ 
%\hspace*{-0.4in}
\includegraphics[width= 3.9cm] {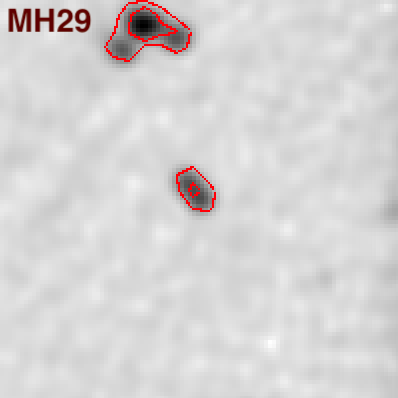} 
\includegraphics[width= 3.9cm] {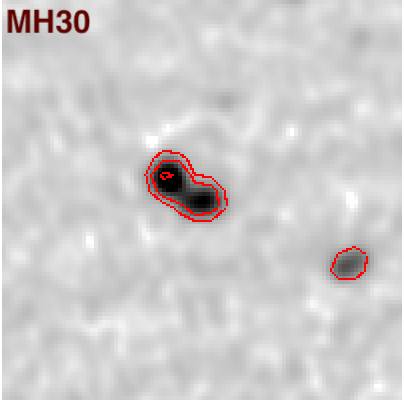}
\includegraphics[width= 3.9cm] {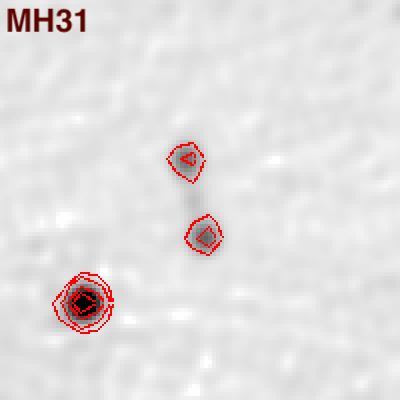}
\includegraphics[width= 3.9cm] {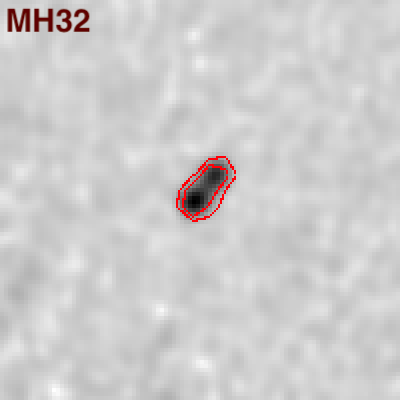}\\
%\hspace*{-0.4in}
\includegraphics[width= 3.9cm] {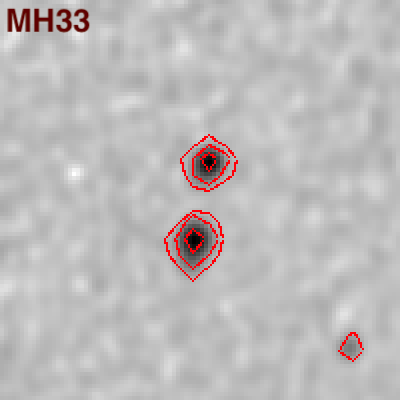} 
\includegraphics[width= 3.9cm] {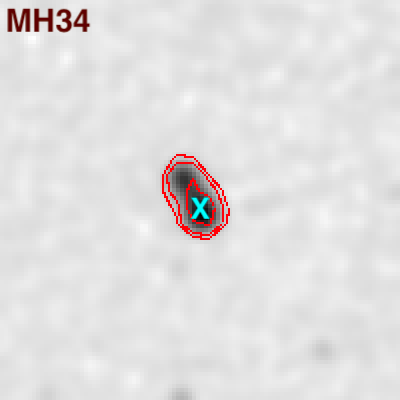} 
\includegraphics[width= 3.9cm] {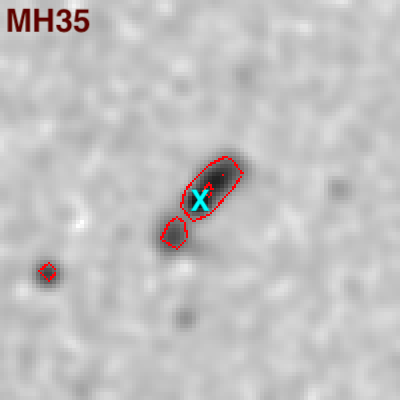} 
\includegraphics[width= 3.9cm] {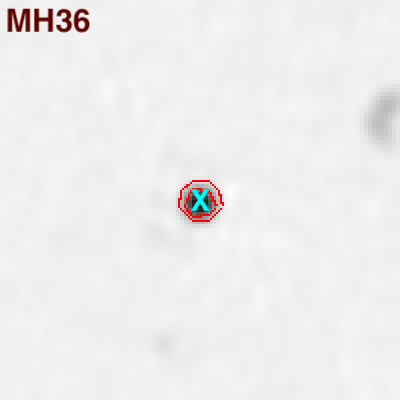}\\

%\vspace*{-0.1in}
\includegraphics[width= 3.9cm] {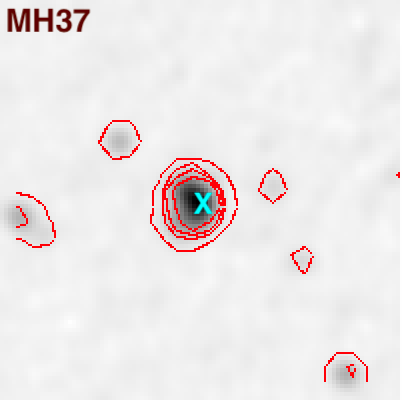} 
\includegraphics[width= 3.9cm] {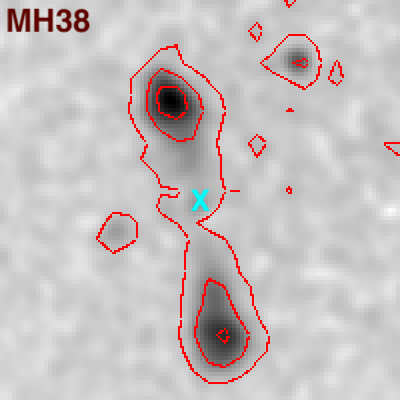} 
\includegraphics[width= 3.9cm] {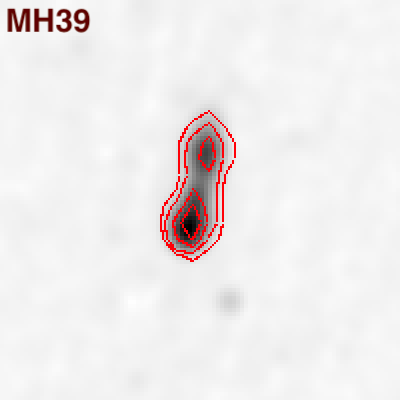}
\includegraphics[width= 3.9cm] {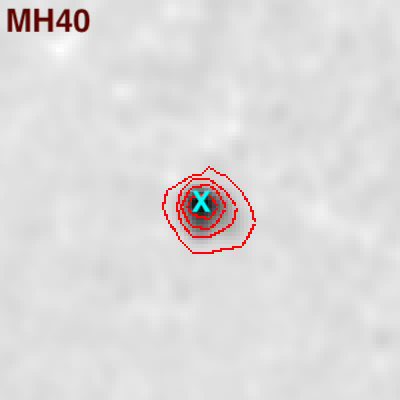}\\ 
%\vspace*{-0.1in}
\includegraphics[width= 3.9cm] {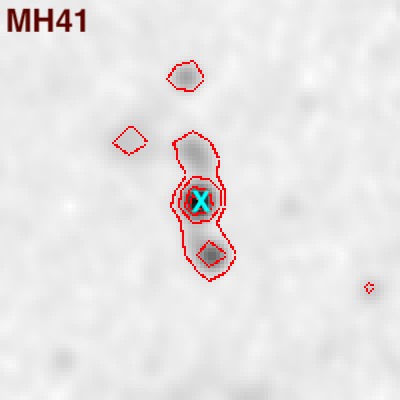} 
\includegraphics[width= 3.9cm] {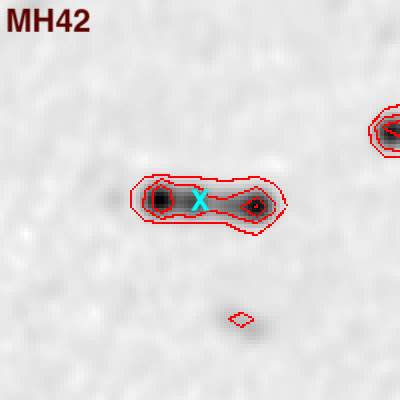} 
\includegraphics[width= 3.9cm] {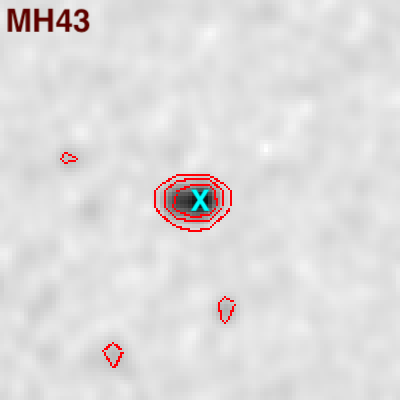}
\includegraphics[width= 3.9cm] {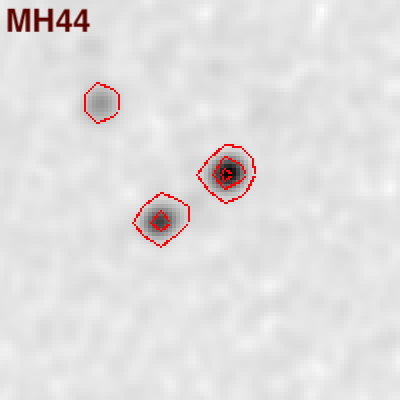}\\ 
\caption{List of additional resolved galaxies similar to Figure \ref{fig4}. Each stamp is 5$\times$5 arcmin and the beam size is 13.5 arcsec. See Table\,\ref{table4} for measurements. The contours are 20\,$\mu$Jy, 50\,$\mu$Jy, 150\,$\mu$Jy and 300\,$\mu$Jy. The cyan crosses represent the positions of the WISE's  galaxy catalogue counterparts where detected. }  \label{fig6.2b}
\end{figure*}

Figure \ref{fig6.2b} presents more resolved galaxies from our radio sample in addition to Figure \ref{fig4}. Their radio fluxes are in Table \ref{table4}.

\begin{figure*}[!h]
\centering
\includegraphics[width= 14cm] {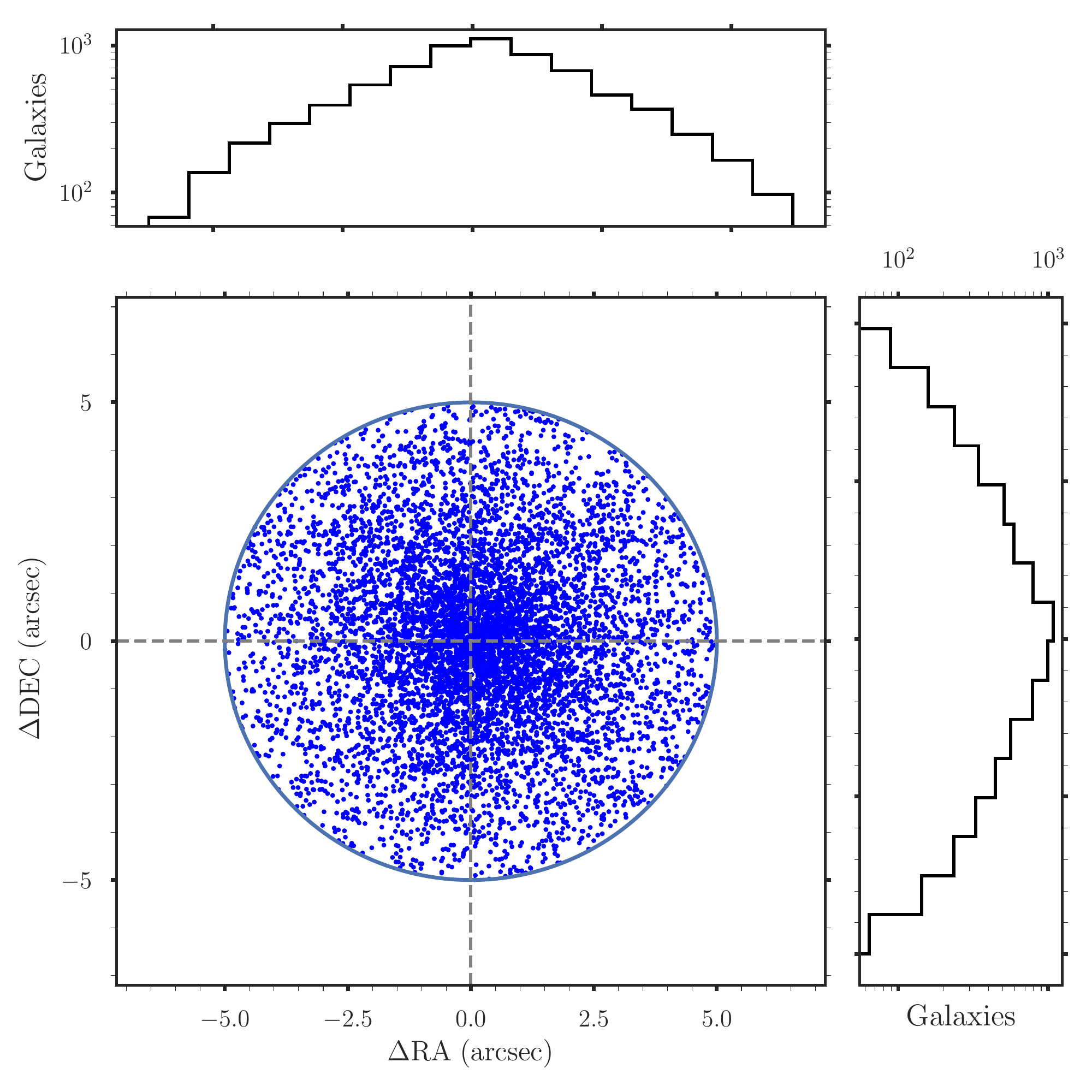}
\caption{Position offset between MeerHOGS\,(M) and $\textit{WISE}$\,(W). The average\,(median) RA\,(M)\,-\,RA\,(W) = 0.1268\,(0.1441) and  Dec\,(M)\,-\,Dec\,(W) = 0.0202\,(0.0305) arcsec,  which confirms good astrometry for the MeerKAT instruments. } \label{fig6.1b}
\end{figure*}

Figure \ref{fig6.1b} shows the difference between MeerHOGS and $\textit{WISE}$ pointing. The offset are minimal with  average values of  0.1268 and  0.0202 arcsec along RA and DEC, respectively.

\newpage
\vspace*{2in}

\newpage

\begin{table*}[!h]
 \caption{Radio flux densities of the components of the resolved galaxies selected in Fig\,\ref{fig6.2b}. The asterisks indicate galaxies with GAMA spectroscopic redshifts.} \label{table4}
%\centering
\hspace*{-0.2in}
\small{
\begin{tabular}{lccccccc}
\hline  \hline
Galaxy id.   &      RA        & Dec                &  $\mathrm{N. comp.}$               &  $\mathrm{Left. comp.}$             &  $\mathrm{central. comp.}$  &  $\mathrm{Right. comp.}$ &  $\mathrm{Total\;flux}$\\
                   &            (deg)    &      (deg)              &                                 &        (mJy)                       & (mJy)                    &      (mJy)               &(mJy)           \\   \hline
  MH21  &         345.5588  &     -32.16678        &       2                        &   0.4              &   N/A              &   0.06                &   0.46   \\
  MH22  &       345.84269  &    -32.89788         &   2                            &   14.03              &   N/A              &   15.85                &   29.88   \\
MH23  &       345.0538   &    -33.0967            &   2                              &   2.67              &  N/A              &   6.69                &   9.35   \\
MH24  &        344.252    &    -33.37644          &   2                              &   N/A              &   10.62             &   N/A                 &   10.62   \\
MH25 &         343.06187 &     -33.92648       &   2                               &   3.72              &  N/A              &   7.1                &   10.82   \\
 MH26 &         343.21541 &     -34.31053       &   2                              &   5.12              &  N/A              &   12.1                &   17.21   \\
 MH27 &         342.74987 &     -34.36203       &   2                              &   4.01              &   N/A              &   3.06                &   7.07   \\
   MH28 &         344.07075 &     -34.76574       &   2                            &   5.54              &   N/A              &   6.02                &   11.56   \\
   MH29 &         341.3357  &     -34.77133        &   3                            &   N/A               &   3.46             &   N/A                 &   3.46   \\
   MH30 &         342.6264  &     -35.2621          &   2                            &   8.9              &  N/A              &   5.92                &   14.81   \\
   MH31 &         341.13076 &     -35.21164       &   3                            &   2.88              &   0.48             &   3.12                &   6.49   \\
   MH32 &         342.01801 &     -35.4946        &   2                             &   5.16              &  N/A              &   3.83                &   8.99   \\
   MH33 &        345.80901 &     -31.67244       &   2                             &   8.87              &  N/A             &   12.35                &   21.22   \\
  MH34      &   345.16548    &   -31.96997        &        2                       &   9.58              &   N/A              &   13.79                &   23.38   \\
  MH35      &   345.76627    &   -32.54131        &        3                       &   2.07              &   2.41             &   3.52                &   8.0   \\
  MH36 *     &   343.76173    &   -33.30183        &        1                     &   N/A              &   40.22             &  N/A                 &   40.22   \\
  MH37      &   344.39038    &   -33.76065        &        1                       &  N/A               &   30.65             &   N/A                 &   30.65   \\
  MH38      &   342.81942    &   -33.82685        &        3                       &   12.03              &   0.88             &   10.98                &   23.89   \\
  MH39      &   342.83231    &   -34.35561        &        2                       &   76.15              &   N/A              &   45.95                &   122.09   \\
  MH40      &   341.65759    &   -34.43136        &        1                       &   N/A               &   22.44             &   N/A                 &   22.44   \\
  MH41  *    &   341.85518    &   -34.60437        &        3                     &   3.52              &   10.75             &   3.88                &   18.16   \\
  MH42      &   343.93365    &   -34.68008        &        3                       &   12.98              &   9.77             &   15.07                &   37.82   \\
  MH43      &   342.72356    &   -34.88700        &        1                       &   N/A              &   7.59             &   N/A                 &   7.59   \\
  MH44      &   343.73528    &   -33.10917        &        2                       &   8.63              &   N/A              &   14.68                &   23.31   \\  \hline

\end{tabular}
}
\end{table*}

\newpage

\section{Optical-Infrared Rest-Frame Uncertainties in the W3 band}    %%%%%%%%%%%%%%%%%%%%%%%%%%%%%%%%%%%%%%

In the optical-infrared window, correcting for the (1+z) photometric band-shifting is not as simple as, e.g. applying a power-law scaling correction (as with the radio continuum; e.g. see Eq. 2). This is because of the varying contributions from the stellar continuum, emission lines and ISM components (e.g. dust) that dominate in this wavelength range for the range of galaxies that are detected and characterized. The technique adopted in this study, developed by the WXSC, uses a set of templates that are compared to the optical-infrared (0.4 to 22\micron) photometric SED measurements, thereby identifying the best fit template that is then used to scale the ``observed" measurements to the equivalent ``rest" measurements.  The list of composite templates we deploy are from  Brown et al. (\citeyear{Brown2014}) and Spitzer-SWIRE/GRASIL (Polletta et al.
 \citeyear{Polletta2006},  Polletta et al. \citeyear{Polletta2007},  Silva et al. \citeyear{Silva1998}),
 totaling over 135 composite templates that range from early-type spheroidals to late-type spirals, and various AGN types.  Further details of the SED template fitting are given in   Jarrett et al.  (\citeyear{Jarrett2013}, \citeyear{Jarrett2017}).

\begin{figure*}[!thb]
\begin{center}

\includegraphics[width=0.49\textwidth]{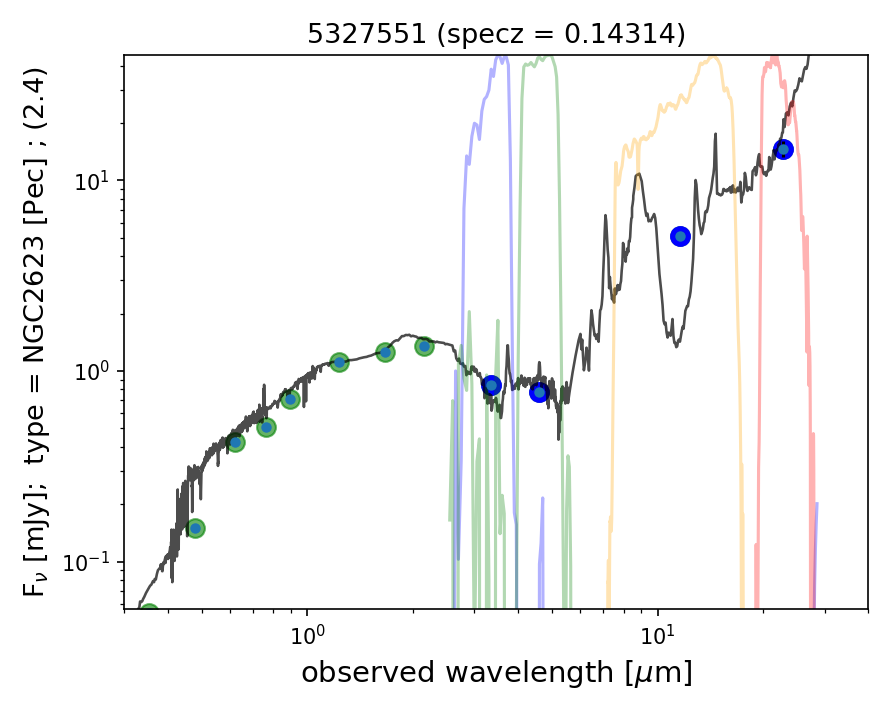}
\includegraphics[width=0.49\textwidth]{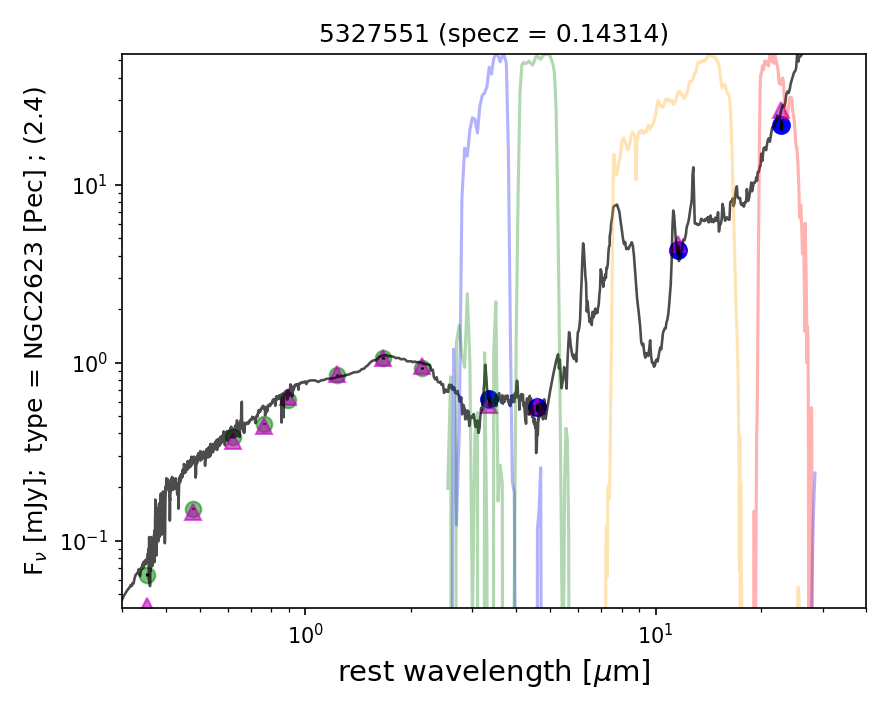}
\vspace{-10pt}
\caption{Spectral energy distribution (SED) of the GAMA source 5327551, whose redshift is 0.14314.   The optical and near-IR data are from the GAMA-G23 survey (green  points), and the mid-IR from the WISE-WXSC (blue points). The best fit ($\chi\,=2.4$) template (black line) was found to be that of NGC\,2623 \cite[from][]{Brown2014}.
LEFT panel shows the observed fluxes from GAMA-G23 (optical bands, green points) and WISE-WXSC (mid-infrared, blue points).  LEFT panel shows the observed (i.e., redshifted) fluxes, and RIGHT panel the k-corrected rest-frame fluxes.  Also shown are the WISE filter traces (scaled to unity; blue, green, orange and red lines), as given in \citet[]{Jarrett2011}.  
%This example would be considered a good $\chi^2$ fit. Note that the W3 k-correction is negative ("rest" is brighter) while for W4 it is positive ("rest" is considerably fainter because of the steeply rising continuum).  
}
\label{fig:SED}
\end{center}
\vspace{-15pt}
\end{figure*}

In this appendix, we assess the uncertainties that arise from this rest-frame correction process.  Potential sources of scatter in the rest-frame flux derivation are:  (1) photometric uncertainty in the observed fluxes, (2) number of bands that are used in the fitting process, (3) the weighting of the bands, (4) the finite number of templates and galaxy-type coverage, and (5) the dominant features in the galaxy spectrum.   An example of the complexity of determining robust rest-frame corrections is shown in Fig~\ref{fig:SED}, the optical-infrared SED of an active star-forming galaxy. The figure shows the before -- observed fluxes and redshifted template, and after -- k-corrected rest-frame fluxes with the best-fit template. This example would be considered a good $\chi^2$ fit (with values less the 3); the NGC\,2623 composite template is well-matched (although not perfect) to the actual photometric measurements for this GAMA-G23 source (CATAID: 5327551). Note that the W3 k-correction is negative, i.e. the ``rest" flux is fainter than the observed flux because the bright 7.7\micron PAH feature shifts out of the W3 band. Conversely, for W4 the correction is positive, i.e., ``rest" is considerably brighter because of the steeply rising warm-dust continuum that $\lambda$-shifts longward.  

The template fitting, using inverse variance weighting,  gives higher-weight to the near-infrared fulcrum from 1-5\micron\ because it is sensitive to the bright, evolved luminous populations that are seen in nearly all galaxy types. Since our photometric bands are wide, we are never sure about the narrow in-band features, such as the depth of the $[$Si$]$ absorption features or the strength of the PAHs, for example. We are unable to estimate uncertainties due to such in-band features, but we are able to track the rest-frame repeatability that our set of templates allow.

\begin{figure*}[!thb]
\begin{center}
\includegraphics[width=0.49\textwidth]{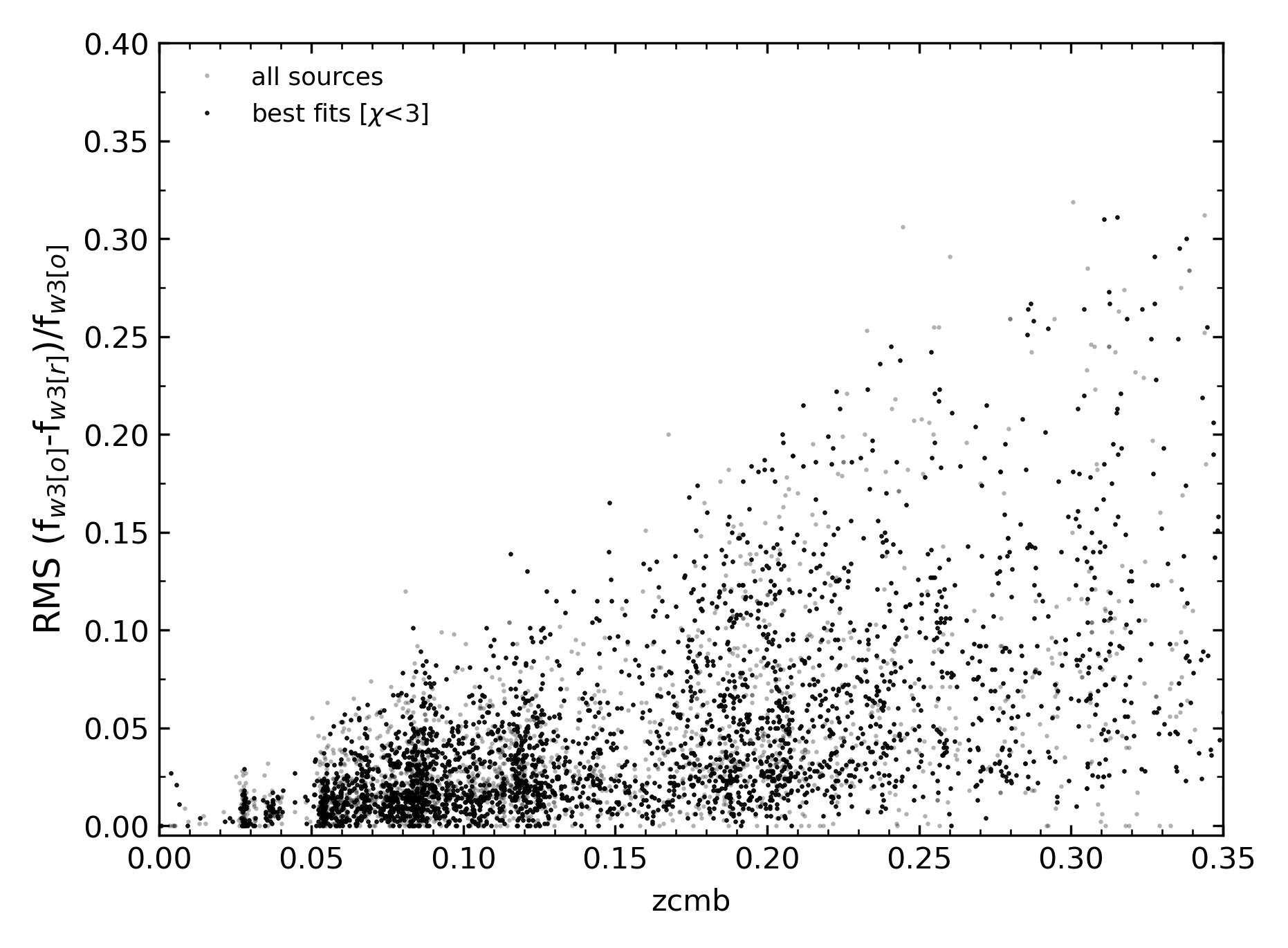}
\includegraphics[width=0.49\textwidth]{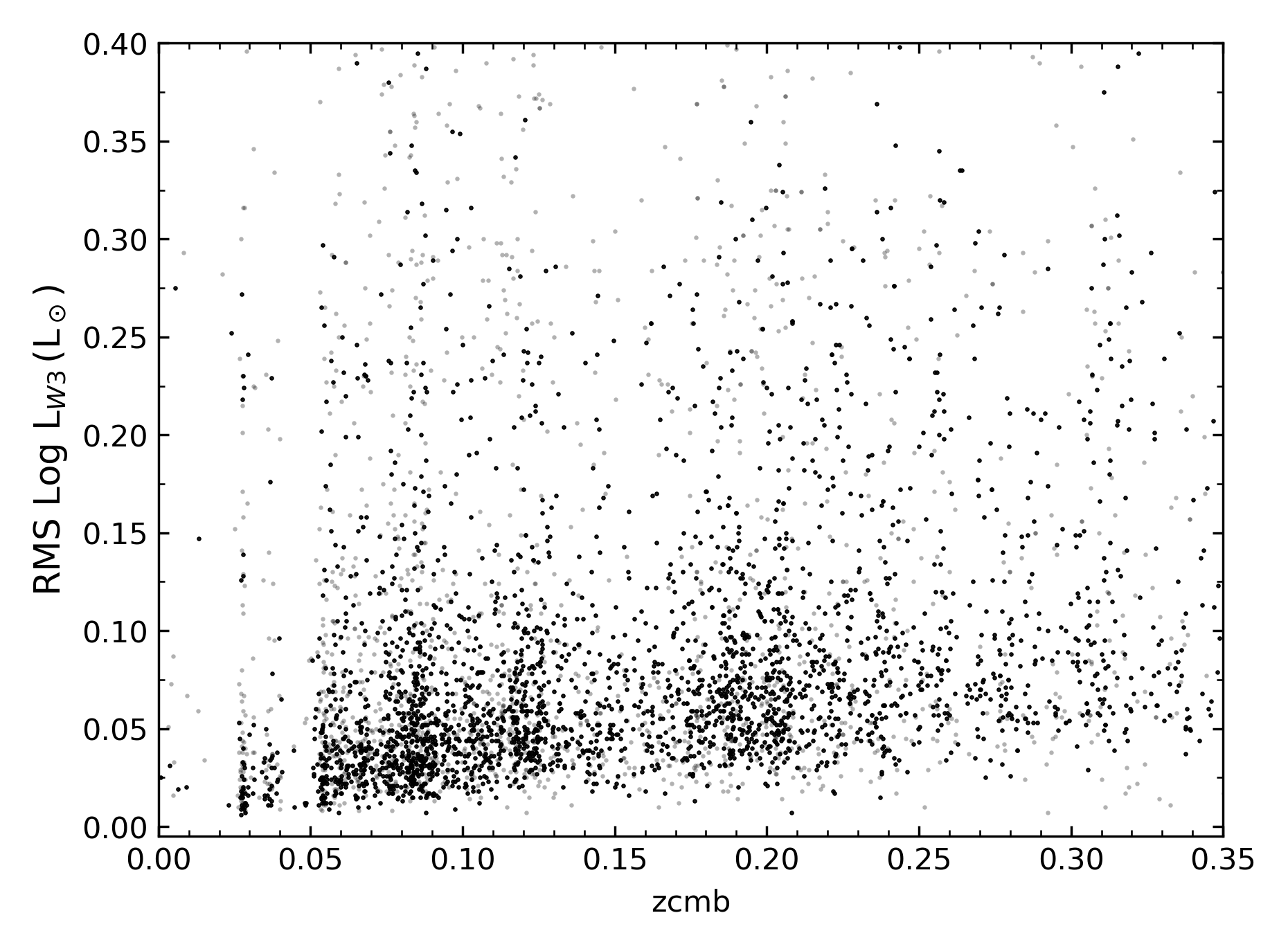}
\vspace{-10pt}
\caption{Statistical results showing the resulting W3 uncertainty in the rest-frame correction method using random scattering in the input photometric measurements that comprise the optical-infrared SED for each galaxy.  LEFT panel is the scatter in the comparison between the observed W3 flux and the rest flux, normalized by the mean observed flux.  For low redshift, most galaxies have $<$10\% scatter, increasing to 15-20\% for the highest redshifts of this study.    RIGHT panel is the resulting total scatter seen in the W3 luminosity (Log L$_{W3}$, used to compute the SFR),  which increases with redshift, but is typically less then 10-15\% for most galaxies.   Large deviations ($>$20\%) are from poor template fits.    
 }
\label{fig:Scatter}
\vspace{-15pt}
\end{center}
\end{figure*}

For this we run a set of Monte-Carlo random sampling trials in which slight changes to input photometry and templates may result in rest-frame fluxes that differ across trials. In each trial, we adjust the observed (input) fluxes based on their photometric error (Gaussian sigma), find the best fit template and carry out the rest-frame corrections, and then recompute the ``rest" fluxes and corresponding luminosities.  For the purposes of this study, where the SFR is derived from the W3 (12\micron) measurements, we track the RMS scatter in the W3 (12\micron) flux and corresponding luminosity, L$_{W3}$.  Results are shown in Fig~\ref{fig:Scatter}, described below.

 By allowing the input fluxes to vary based on their error estimates, the best fit template may change from the original, or may remain the same, but the W3 rest flux is guaranteed to change in either case since the input changed due to randomly sampling of its error model.  Here we are interested not in the photometric uncertainty (which is already well characterized, and propagated into our luminosity errors, etc.),  but rather how the rest flux changes relative to the input observed flux in order to gauge the template fitting uncertainties.
 
\noindent Through the set of sampling trials, the total \% scatter in the W3 rest vs. observed fluxes is shown in Fig~\ref{fig:Scatter}a, where each point is a galaxy where we have measured the RMS scatter in the W3 $[$observed$-$rest$]$ flux normalized by the mean observed (input) flux in the sampling distribution. This scatter, which represents the stochasticism of the template-fitting itself, ranges from a few \%  to typically less than 10-15\% across the full redshift range of our study.  There is a clear trend with redshift, likely driven by sensitivity (higher redshift galaxies tend to be fainter, but also tend to be higher luminosity, including AGNs) and mis-matched template scaling errors.  This translates to a scatter in the W3 luminosity, shown in Fig~\ref{fig:Scatter}b,  also increasing with redshift.   The total Log L$_{W3}$ scatter is typically less then 5\% for nearby galaxies (z$<$0.1), and ranges upwards to 10\% for higher redshifts, but may be much higher (20 to 40\%) for outliers that arise from poor template fits, either mismatched band-to-band measurements (i.e., poor colors) or simply there is no appropriate template for the galaxy.  These results are encouraging: most galaxies should have only a small, $<$\,5-10\% uncertainty that arises from our optical-infrared rest-frame correction method. However, it should be noted that considerably larger errors may be induced by poor template fits. 

\end{document}